\documentclass[%
 aip,
 amsmath,amssymb,
 reprint,%
]{revtex4-1}

\usepackage{mathtools}
\usepackage{hyperref}
\usepackage{xcolor}

\usepackage{graphicx}
\usepackage{dcolumn}
\usepackage{bm}

\usepackage[utf8]{inputenc}
\usepackage[T1]{fontenc}
\usepackage{mathptmx}

\DeclareRobustCommand{\v}[1]{
  \ifcat#1\relax
    \boldsymbol{#1}
  \else
    \mathbf{#1}
  \fi}

\begin{document}

\preprint{AIP/123-QED}

\title[Using machine-learning modelling to understand macroscopic dynamics in a system of coupled maps]{}
%


\author{Francesco Borra}
 \email{francesco.borra@uniroma1.it.}
\author{Marco Baldovin}
 \email{marco.baldovin@infn.roma1.it.}
\affiliation{ 
Dipartimento di Fisica, Sapienza Università di Roma, p.le A. Moro 5, I-00185, Roma, Italy
}%

\date{\today}

\begin{abstract}
Machine learning techniques not only offer efficient tools for modelling 
dynamical systems from data, but can also be employed as frontline investigative 
instruments for the underlying physics. Nontrivial information about the 
original dynamics, which would otherwise require sophisticated ad-hoc 
techniques, can be obtained by a careful usage of such methods. To illustrate this 
point, we consider as a case study the macroscopic motion emerging from a system
of globally coupled maps. We build
a coarse-grained Markov process for the macroscopic dynamics both with a machine 
learning  approach and with a direct numerical computation of the transition 
probability of the coarse-grained process, and we compare the outcomes of the 
two analyses. Our purpose is twofold: on the one hand, we want to test the 
ability of the stochastic machine learning approach to describe nontrivial 
evolution laws, as the one considered in our study; on the other hand, we aim at 
gaining some insight into the physics of the macroscopic dynamics 
by modulating the information available to the network, we are able to infer 
important information about the effective dimension of the attractor, the 
persistence of memory effects and the multi-scale structure of the dynamics.
\end{abstract}

\maketitle

\begin{quotation}
We show how simple machine learning techniques can be employed as an 
investigation tool for studying physical properties of generic dynamical systems from 
data. We examine, as a case study, the macroscopic dynamics of a large ensemble 
of chaotic coupled maps. Due to a known violation of the law of large numbers, 
the considered system displays a non-trivial collective behaviour, which cannot 
be simply modelled by standard coarse-graining techniques.

We show that, even in this case, macroscopic dynamics can be 
accurately reconstructed from data by building a Markovian coarse-grained model 
via neural networks. The results can be compared with those of a more direct 
approach, based on the reconstruction of transition probabilities. We highlight 
the importance of a stochastic effective description. 

Moreover, a careful control of the variables 
actually employed by the network in the modelling process allows to extract 
important information about the system itself. Realistic upper bounds to the dimension
of the macroscopic attractor can be inferred, and the occurrence of memory effects
can be studied in some detail.

\end{quotation}

\section{\label{sec:intro}Introduction}

``Coupled maps'' is an umbrella term for high dimensional systems composed of a 
large number of interacting units, whose features make them useful toy models 
related to many physical domains, including neural networks, turbulence and 
biology~\cite{kaneko1993theory,kaneko1992overview,bohr2005dynamical}. While the collective behaviour of many high dimensional 
systems is trivial, due to self averaging properties, relatively simple map 
structures display instead a complex emergent phenomenology.
Maps on lattice even exhibit pattern formation~\cite{kaneko1985spatiotemporal,ouchi2000coupled}, but also non-structured 
globally coupled map-systems can produce a non-trivial macroscopic behaviour, which 
can be detected with suitable intensive observables. The underlying reason is that properly designed interactions 
induce synchronisation between units and result in the formation of coherent 
collective structures. These macroscopic dynamics have been shown to 
exhibit a variety of behaviours, including chaos~\cite{takeuchi2011extensive},
and have been studied with different approaches, such as
macroscopic perturbations or finite size Lyapunov exponents (FSLEs)~\cite{aurell1996growth,boffetta1998extension,shibata98,cencini99},
and with the analysis of the 
Frobenius-Perron equation~\cite{takeuchi2009lyapunov,pikovsky1994globally}. Some work show that
that the emergence of slow or collective dynamics can be 
understood in terms of existence of a low dimensional effective manifold, whose 
presence can be detected by identifying covariant Lyapunov vectors~\cite{carlu2018lyapunov,froyland2013computing,takeuchi2009lyapunov,
ginelli2013covariant} involving an extensive number of microscopic degrees of 
freedom~\cite{takeuchi2011extensive}. This method even allows, in some cases, 
to estimate the effective dimension of the collective dynamics~\cite{yang2009hyperbolicity}.

While the possibility of deriving some macroscopic properties from microscopic ones
is crucial to the analysis of the 
global behaviour of the system, also purely macroscopic dynamics reconstruction from
macroscopic data is relevant for a number of reasons. Indeed Lyapunov analysis is 
computationally expensive for large systems, and it requires excellent knowledge 
of the microscopic structure, so that its application may result unfeasible in 
many cases. Most important, in many realistic physical scenarios one has only 
access to global observables: one is then interested in building a 
low-dimensional, coarse-grained model from data, with the purpose of making 
reliable predictions at the macroscopic level. In the coupled maps framework, 
this problem involves features which are common to several physical problems, 
such as turbulence or climatology: the low-dimensional collective behaviour, the 
multiscale structure, the difficulty to distinguish high dimensional 
deterministic dynamics from noise. Many approaches have been developed to face 
this kind of problem, most of which are based on the assumption that a 
relatively simple stochastic differential equation properly describes the 
dynamics. The coefficients can be found by mean of a careful analysis of 
conditioned moments~\cite{siegert1998analysis,friedrich2011approaching}, even in 
the case of memory kernels~\cite{lei2016data}, or via a Bayesian 
approach~\cite{ferretti2020building}. This kind of strategy has been 
successfully employed in many fields of physics, including turbulence, soft 
matter and biophysics~\cite{friedrich2011approaching,peinke2019fokker, 
baldovin2019langevin, bruckner2020inferring}.
In recent years many different machine-learning approaches were also developed.
While even pure model-free methods can be very efficient~\cite{lu2017reservoir,ottPRL,borra2020effective},
other approaches aim to blend physical information with data-only techniques~\cite{wikner2020combining,ottHybrid,Lei20}. An important role is played by those
machine learning methods attempting to extract an effective low dimensional 
dynamics, for instance by employing autoencoder based networks~\cite{pathak2020using,linot2020deep}.

In this paper we propose a general data-driven approach to macroscopic dynamics emerging from globally coupled maps. We show that it is possible to build coarse-grained 
stochastic models for the collective signal generated by the maps
using a stochastic machine-learning method,
which allows to account both for the existence of a low-dimensional
effective manifold and for the presence of residual
high-dimensional dynamics. This stochastic approach is in line
with theoretical arguments that stochastic modelling is the correct
coarse graining procedure in hydrodynamical-like system~\cite{palmer2014real,palmer2019stochastic,eyink2020renormalization}. The 
results of our analysis are carefully compared to those achievable by
mean of a direct numerical computation of transition probabilities.
Remarkably, such approach to model reconstruction also yields 
relevant physical information about the underlying physical process, implying 
that the quality of forecasts and the physical insight are strongly 
interconnected. We therefore explore the possibility to overcome  
the separation between a purely result-oriented approach and 
theoretical investigation, which is a promising direction for machine learning
based approaches~\cite{corbetta2019deep,buzzicotti2020reconstruction,beintema2020controlling,linot2020deep}.

 The paper is organized as follows. In Section~\ref{sec:model} we
describe a model of coupled maps, which will be used to test our proposed
data-driven analysis. Section~\ref{sec:dynamics} is 
devoted to the description of two alternative methods to infer a stochastic 
dynamics from data: first, we employ a standard frequentistic 
approach to build a Markov process in which the state of the system is given by 
the variable and its discretized time-derivative; then, we discuss a stochastic
machine-learning method, designed to assign to a given portion of 
trajectory the probability distribution for the next value of the observed 
variable. Implementation details for the two methods are described in Section~\ref{sec:setting},
which is devoted to the discussion of validity conditions for the previously introduced strategies.
The outcomes of the two approaches and their ability to mimic the 
original dynamics are discussed and compared in Section~\ref{sec:discussion}. In 
Section~\ref{sec:conclusions} we draw our conclusions.

\section{\label{sec:model} Macroscopic motion in globally-coupled maps}

 One of the simplest and most investigated models of coupled maps, introduced by 
Kaneko~\cite{kaneko89}, reads
 \begin{equation}
 \label{eq:dyn}
  y^i_{t+1}=(1-\varepsilon)f_a(y^i_{t})+ \frac{\varepsilon}{N}\sum_{j=1}^N f_a(y^j_{t})\quad  i=1,...,N\,,
 \end{equation}
 where the $N$ variables $\{y^i\}$ belong to the interval $[0,1)$ and $f_a : 
[0,1) \mapsto [0,1)$ is a chaotic map, depending on some constant parameter $a$. 
 Typical choices for $f_a$ include the logistic map, the circle map and the tent 
map~\cite{kaneko90}; in the following we will focus on the last, i.e. 
 \begin{equation}
  f_a(y)=a \left(\frac{1}{2} - \left|\frac{1}{2}-y\right|\right)\,,
 \end{equation}
with $1<a<2$. The character of the discrete-time dynamics~\eqref{eq:dyn} is 
tuned by the coupling parameter $\varepsilon$: in the limit case 
$\varepsilon=0$ all maps evolve independently, while the condition $\varepsilon=1$ sets a 
regime in which the variables are maximally coupled. Intermediate 
values in the range $0<\varepsilon<1$ typically lead to a rich and complex 
phenomenology,  investigated in a series of papers in the 1990s~\cite{kaneko90, 
kaneko92,kaneko95,crisanti1996broken}: synchronisation, violation of the law of large numbers and
tree-structure organization of the chaotic attractors are among the features 
shown by this kind of dynamical systems, due to a trade-off between the 
chaotic properties of $f_a$ and the synchronizing action of the mean field.

 In what follows we will be interested in the behaviour of the mean value
\begin{equation}
 z_t=\frac{1}{N}\sum_{i=1}^N y_t^i\,;
\end{equation}
since $z_t$ is a global observable of the system, from a physical perspective it 
is natural to consider it as a ``macroscopic'' quantity, whereas the $N$ coupled 
maps composing the complete dynamical system~\eqref{eq:dyn} may be regarded as 
the ``microscopic'' variables. The microscopic evolution is  clearly
chaotic as soon as $a(1-\varepsilon)>1$, and it has been noticed~\cite{shibata98,cencini99} that, for suitable 
choices of the parameters $a$ and $\varepsilon$, also the macroscopic dynamics 
may be characterized by a small positive FSLE~\cite{aurell1996growth,boffetta1998extension} (``macroscopic chaos''). While the microscopic attractor dimension is typically  extensive in $N$, 
the macroscopic attractor has a low dimension, almost independent of $N$.

\begin{figure}
\includegraphics[width=0.99\linewidth]{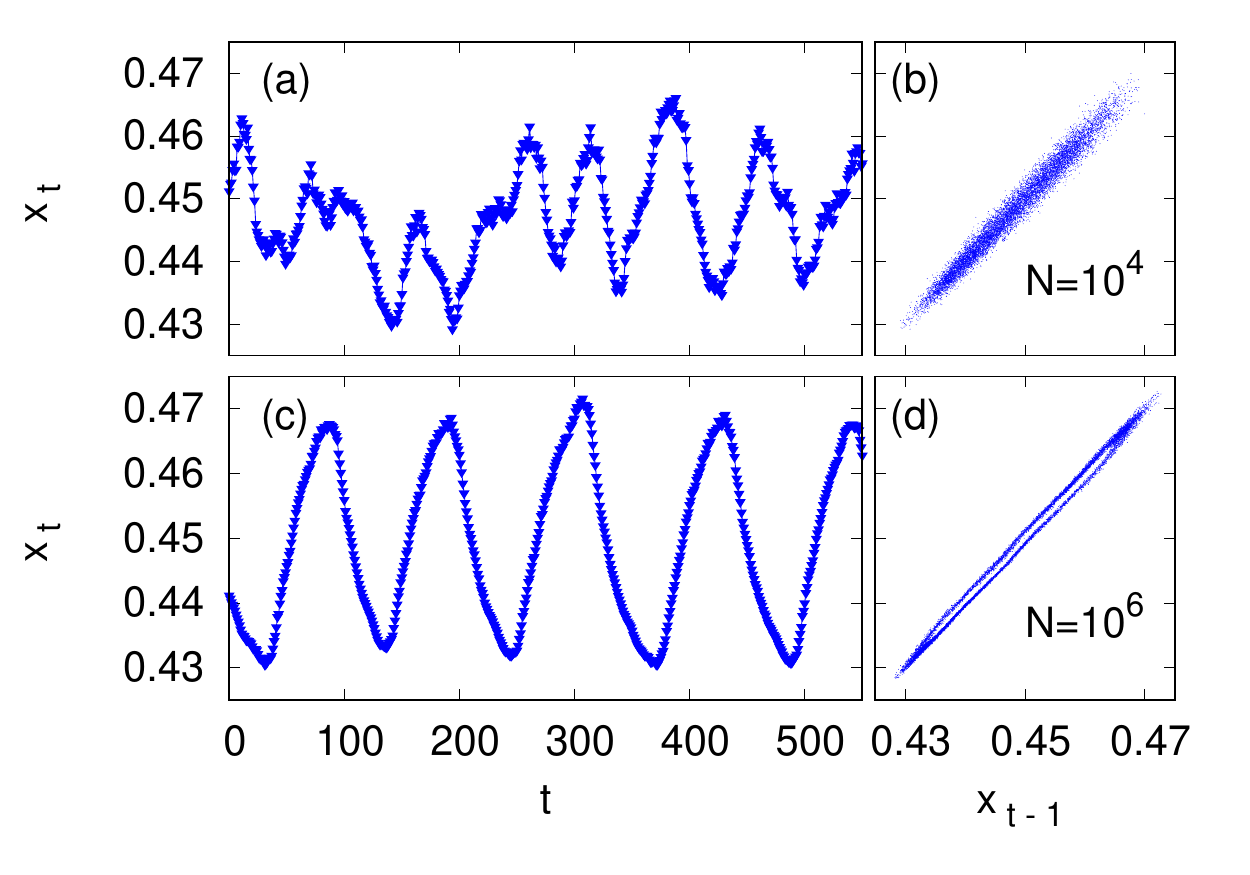}
 \caption{\label{fig1} Dynamics of $x_t$. For two different choices of the 
number of maps (top $N=10^4$, bottom $N=10^6$), sample trajectory of $x_t$ 
(left), as defined in Eq.~\eqref{eq:xt}, and two-dimensional projection of the 
dynamical attractor (right).}
\end{figure}

\begin{figure}
\includegraphics[width=0.99\linewidth]{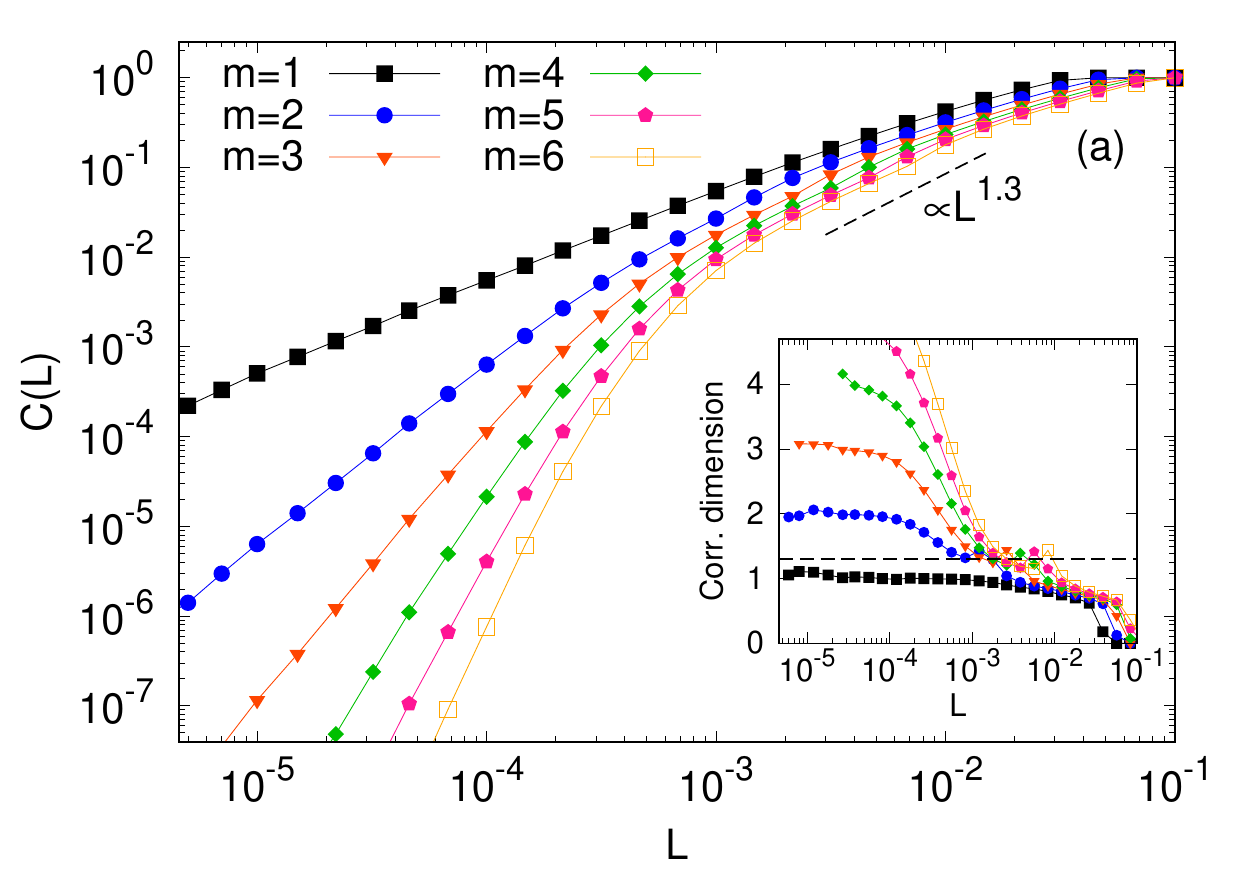}
\includegraphics[width=0.99\linewidth]{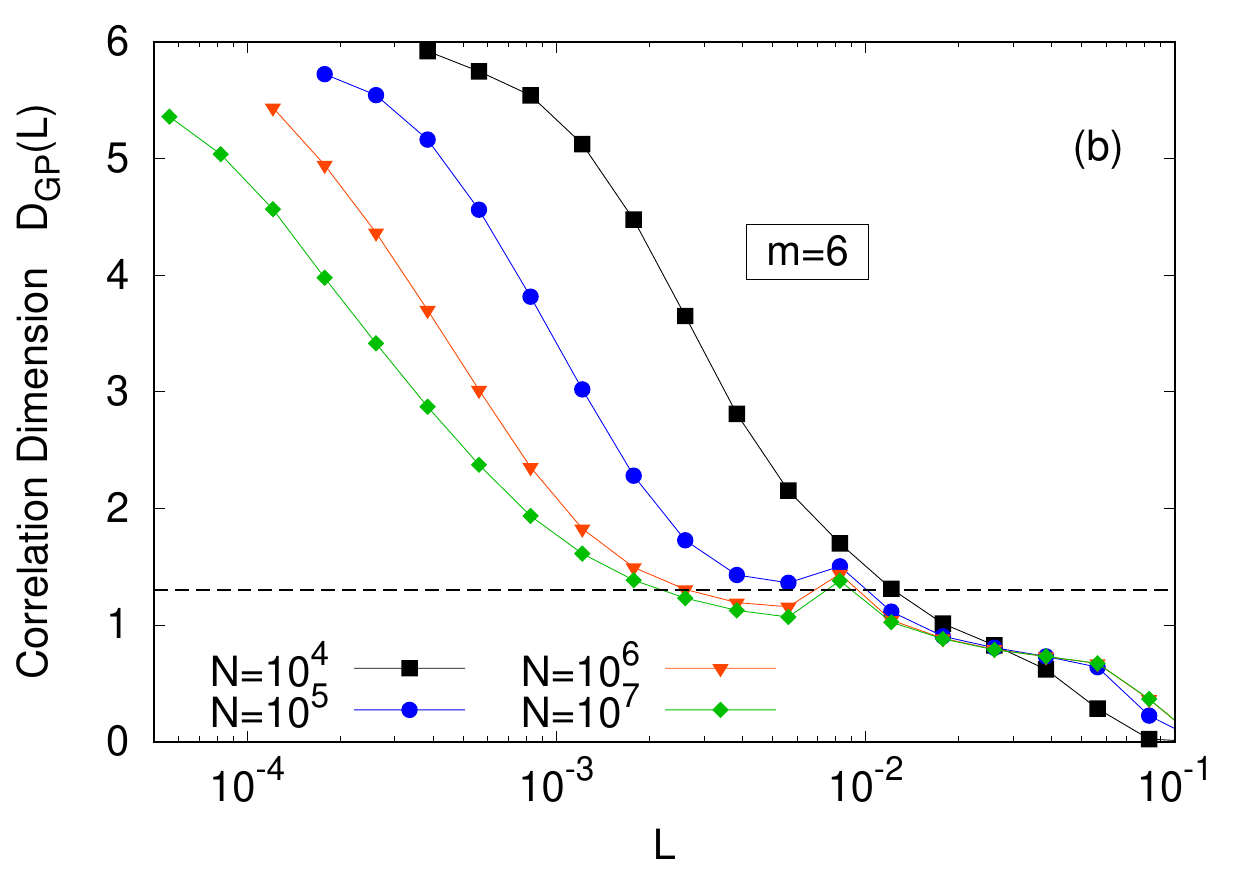}
 \caption{\label{fig2} Grassberger-Procaccia analysis. Panel (a): correlation integral defined by Eq~\eqref{eq:gp_discr}, for different values of the embedding dimension $m$. Here $N=10^6$. In the inset, correlation dimension defined by Eq.~\eqref{eq:corrdim} as a function of $L$; the dashed line is a guide for the eyes, corresponding to dimension 1.3. Panel (b): dependence of 
 $D_{GP}(L)$ 
 on the number of maps $N$, for 
  $m=6$.}
\end{figure}

In the following we will consider the case represented by the couple
\begin{equation}
\label{eq:parameters}
a=1.7 \quad \varepsilon=0.3\,,
\end{equation} 
investigated by Cencini et al.~\cite{cencini99}. In this example the trajectory of $z_t$ is characterized by a 4-steps dynamics, 
whose origin is related to the two-band structure of the maps distribution at a given time~\cite{kaneko95}, discussed in some detail in Appendix~\ref{sec:appendixA}. It 
is found that after 4 time-steps $z_t$ gets very close to its former value, so 
that the quantity \begin{equation}
\label{eq:xt}
 x_t:=z_{\,4t}\,,
\end{equation}
might be well approximated by some unknown continuous process (possibly 
stochastic). The dynamics of $x_t$ crucially depends on the number of maps 
composing the system; this aspect is clearly evident from Fig.~\ref{fig1}, 
where a branch of the trajectory is reproduced for two choices of $N$. For 
higher values of $N$ the trajectories tend to an increasingly regular, 
almost-oscillatory behaviour, with period $\simeq 10^2$: this periodicity may 
suggest a second-order structure, a possibility also supported by the shape of 
the two-dimensional projection of the dynamical attractor in the $N=10^6$ case 
(Fig.~\ref{fig1}, bottom right panel).
The standard Grassberger-Procaccia correlation 
analysis~\cite{grassberger1983characterization,vulpiani2010chaos} provides
a better insight into the dimensionality of this ``macroscopic'' 
dynamics. Let us 
consider the $m$-dimensional embedding vector (also called ``delay vector'')
\begin{equation}
\label{eq:embvect}
 \v{x}^{(m)}_t=(x_t,x_{t-1},...,x_{t-m+1})^T\,,
\end{equation} 
and define the natural measure $\mu: \mathbb{R}^m \mapsto \mathbb{R}$ 
corresponding to the evolution of $x_t$. The correlation integral
\begin{equation}
 \label{eq:gp}
 G_m(L)=\int d\mu(\v{u}^{(m)})\, d\mu(\v{w}^{(m)}) \,\mathbb{H}(L-|\v{u}^{(m)}-\v{w}^{(m)}|)\,,
\end{equation}
where $\mathbb{H}$ stands for the Heavyside step-function and $|\cdot|$ is the 
Euclidean norm, can be measured in numerical simulations with a proper sampling 
on the dynamical attractor:
\begin{equation}
 \label{eq:gp_discr}
 G_m(L)\simeq \frac{2}{T(T-1)}\sum_{t=1}^{T-1} \sum_{s=i+1}^{T} \mathbb{H}(L-|\v{x}^{(m)}_t-\v{x}^{(m)}_s|)\,,
\end{equation} 
$T$ being the length of the considered trajectory.
The exponent of the power law which better 
approximates the local behaviour of $G_m(L)$ identifies the local correlation dimension 
of the dynamic attractor, clearly bounded by $m$, at the considered length-scale 
$L$ . In other words, we can define the correlation dimension
\begin{equation}
\label{eq:corrdim}
D_{GP}(L)=\frac{d [\ln C]}{d [\ln L]}\,,
\end{equation} 
which gives a measure of the dimension of the attractor at the scale $L$. 
Fig.~\ref{fig2}a suggests that $D_{GP}$ is characterized by (at least) two 
different regimes: for small values of $L$ we observe $C_m(L) \sim 
L^m$, i.e.  $D_{GP}$, in each curve, tends to its maximum value $m$ (smaller than the attractor dimension); 
conversely, at larger length-scales all curves with $m>1$ reach a slope 
$\sim 1.3$ in the log-log plot, meaning that at macroscopic scales the attractor 
has a low-dimensional structure. Let us also stress that the cross-over length $L_{crossover}$ 
between the two regimes depends on the number of maps $N$, as shown in 
Fig.~\ref{fig2}(b).

In this work we aim at reproducing the nontrivial macroscopic dynamics described 
above by mean of suitable data-driven methods. 
We are interested in a Markovian stochastic description of the process, i.e. we want to determine
the conditional probability
\begin{equation}
\label{eq:pcond}
 p_n(x_{t+1}|x_t,..., {x_{t-n+1}}) = p_n(x_{t+1}|\v{x}^{(n)}_t)
\end{equation} 
for the value assumed by the variable at a generic step, once $n$ previous steps 
are known. As it will discussed in the following Sections, one of the main difficulties in this approach
is represented by the fact that we do not know in advance what value of
$n$ should be considered, and in principle we do not even know whether such value exists. This is a general issue which is typically encountered
when trying to infer a model from data~\cite{baldovin2018role};
in the words of Onsager and
Machlup~\cite{onsager1953fluctuations}: \textit{``How do you know you have taken enough
variables, for it to be Markovian?''} 

In the proposed description, the ``fluctuations'' observed in 
Fig.~\ref{fig2}, originally produced by a deterministic mechanism, result in a
stochastic noise, in line with other probabilistic approaches to deterministic
systems~\cite{pulido2018stochastic,arbabi2019data,boffetta2000predictability}.
The absence of an underlying low-dimensional model for $x_t$ makes 
this task an ideal benchmark for the machine-learning technique we 
introduce in the following Section. Moreover, as discussed in 
Section~\ref{sec:markov}, the oscillatory shape of the dynamics suggests the 
possibility of a second-order like stochastic modelling ($n=2$) based on the direct inference 
of transition probabilities, whose results can be used as a reference to 
evaluate the performance of the machine-learning approach. Let us notice, 
however, that our method may be used, in principle, in a large variety of 
physical situations, its applicability not being specifically related to the particular 
choice of the generating model described above.

\section{\label{sec:dynamics} Stochastic dynamics from data: two alternative approaches}

The starting point of our analysis is the assumption that the trajectory of 
$x_t$ can be fairly described by a stochastic dynamics with a low number of 
variables, determined by the conditional probability~\eqref{eq:pcond}. We will 
only exploit the time-series of actual trajectories, disregarding any additional 
information coming from  the generating $N$-dimensional dynamics: in other 
words, we aim at finding a coarse-graining procedure for the effective 
description of $x_t$, only based on data. In the following we will compare the 
outcomes of two different approaches. The first one, somehow inspired by 
physical intuition, grounds on the assumption that the state of the system can 
be identified by $x_t$ and its discrete time-derivative; the probability of the 
next step, given a certain state, is inferred by a direct inspection of the 
trajectories. The second strategy is based on a machine-learning technique, 
which is able to mimic $p(x_{t+1}|\v{x}^{(n)}_t)$, after a proper training on 
the original data.

\subsection{\label{sec:cost} Validation methods}

Before we introduce our two ways of inferring a Markov process from data, let us 
discuss the validation methods that we will use to evaluate the quality of our 
reconstruction. In general, we will need to test the performance of our methods 
on an independent set of data, the validation set, by using some metrics to test 
the quality of forecasts. The choice of such metrics is not neutral, since, for 
instance, the optimization of short-term predictions and that of long-term 
statistical properties are not equivalent tasks. In particular, even if short 
term errors are very small, the variable could still eventually leave the 
attractor if this is unstable~\cite{boffetta2000predictability}; conversely, 
long term accuracy does necessarily require high short-term resolution. What is 
often done in practice is a proper short-term optimization - we train a model to 
produce short-term forecast - with the implicit actual goal to achieve long-term 
predictions instead. In this paper we follow this approach, but we validate 
the quality of the reconstruction both for short and long times.

A good test for the quality of the reconstruction is the analysis of the Fourier
power spectrum, i.e. of the quantity
\begin{equation}
 S(f)=\Big|\frac{1}{2 \pi T} \sum_{t=0}^{T}e^{ 2 \pi i f t}x_t\Big|^2\,,
\end{equation} 
where $T$ is the length of the considered trajectory. The comparison of the 
spectrum of the  original model to that of the reconstructed dynamics 
constitutes a powerful tool to evaluate the goodness of the latter, at any time 
scale. Since our approaches are based on the optimization of short-term 
predictions, it is reasonable to expect a good matching at high frequencies $f$; 
an agreement also at low $f$ would be a reliable indication that the proposed 
reconstruction has caught the main features of the original model.

To have a neat evaluation of the quality of the method on the short-time 
scales we use the average cross entropy, which is briefly described in 
the following. This choice is be particularly useful in a machine learning 
framework, where it will be used not only as a validation method, but also as a 
metric for the optimization of the neural network. It is also employed in 
Section~\ref{sec:setting} to determine the best setting for both methods.
Assume discrete time and that $x_t$ 
is a continuous-valued dynamical variable. Let 
$p_n(x_t|x_{t-1},....,x_{t-n-1})$ be the true conditional probabilities of the 
process. Conversely, let $q_n(x_t|x_{t-1},....,x_{t-n-1})$ be the conditional 
probability associated to the model dynamics. Bearing in mind the definition
of the delay vector given by Eq.~\eqref{eq:embvect}, we introduce the
\textit{average} conditional cross entropy between the true distribution and the model
as
\begin{equation}
\begin{split} \label{cost_cont}
C_n&=-\int  d\mu(\v{x}^{(n)})\int dx\, p_n(x|\v{x}^{(n)})\;\ln q_n(x|\v{x}^{(n)})\\
&=\langle H[q_n|p_n]\rangle\,,
\end{split}
\end{equation}
where $\mu(\v{x}^{(n)})$ is again the natural measure on the true dynamical attractor. The lower $C_n$, the more similar $p_n$ are $q_n$ are: the minimum of $C_n$ with respect to the possible choices of $q_n$  is achieved by setting $p_n=q_n$. Consistently, the distribution $q_n$ that minimizes $C_n$ also minimizes the average Kullback-Leibler divergence: the two quantities differ by a term not depending on $q_n$:
\begin{align}
\begin{split}
\langle D[q_n\| p_n]\rangle&=-\int  d\mu(\v{x}^{(n)})\int dx\, p_n(x|\v{x}^{(n)})\;\ln \left(\frac{q_n(x|\v{x}^{(n)})}{p_n(x|\v{x}^{(n)})}\right)\\
&=\langle H[q_n|p_n]-H[p_n]\rangle\,.
\end{split}
\end{align}
Note that this metrics only detects discrepancies in one-step forecasts.


If the system is Markovian after $\tilde n$ steps, then the minimum possible value of the average cross entropy is the Shannon entropy rate, given by $q_n=p_n$ for any $n\geq\tilde n$: 
\begin{equation} h=-\int d\mu(\v{x}^{(n)})\int dx\, p_n(x|\v{x}^{(n)})\;\ln p_n(x|\v{x}^{(n)})\mbox{ with }n\geq \tilde n\,.
\end{equation}

If we assume that our model provides an explicit expression for $q_n$, then, given any dataset $\Omega_{test}=\{\v{x}^{(n)}_{(i)},x_{(i)}\}_{i=1}^{T_{test}}$ (the sequence $(\v{x}^{(n)}_{(i)},x_{(i)})$ is a piece of a trajectory consisting of $n+1$ elements), the average cross entropy~\eqref{cost_cont} is the expected value of the following empirical quantity:
\begin{equation}
C_n(\Omega_{test})=\sum_{i=1}^{T_{test}} \log q_n(x_{(i)}|\v{x}^{(n)}_{(i)})\,.
\end{equation}

\subsection{\label{sec:markov}  Position-velocity Markov process (p-vMP): a physics-inspired approach}

The oscillatory behaviour of the $x_t$ trajectory, caught by Fig.~\ref{fig1}, 
clearly reminds of a second-order dynamics. Qualitatively, its time 
evolution may resemble the one generated by an underdamped harmonic oscillator, 
at least for the cases in which $N$ is large. At first sight, one might thus expect
a reasonable approximation of the dynamics to be achieved by mean of simple equations of 
the form
\begin{equation}
\label{eq:langevin}
 \begin{cases}
  x_{t+1}&=x_{t}+v_{t+1}\\
  v_{t+1}&=f(x_{t}, v_{t})+ \sqrt{2 B}\,\eta_t\,,
 \end{cases} 
\end{equation} 
where $f: \mathbb{R}^2 \mapsto \mathbb{R}$ is some smooth function, $B$ is a 
constant (possibly depending on $N$) and $\eta_t$ is a zero-mean Gaussian stochastic process in discrete 
time satisfying $\langle \eta_t \eta_s\rangle=\delta_{t,s}$. In the above 
equation, the variable
\begin{equation}
\label{eq:velocity}
v_t=x_{t}-x_{t-1}
\end{equation}
plays the role of a discrete ``velocity'' for the dynamics of $x_t$, here 
regarded as the position of a particle. If this was the case, under the assumption that $\langle 
v^2\rangle-\langle 
v\rangle^2\ll\langle x^2\rangle-\langle x\rangle^2$ (i.e. the dynamics is close to the ``continuous limit''), one could 
exploit a data-driven approach to derive the form of an approximate discrete-time Langevin 
equation~\cite{friedrich2011approaching,peinke2019fokker, baldovin2019langevin, baldovin2020effective};
this strategy is based on the possibility to infer an 
approximate functional form for $f(x,v)$ by considering the small-time limit of 
suitable conditioned moments. Unfortunately, the model inferred with this method (not shown)
is not able to reproduce the original trajectory, a clear hint that 
Eq.~\eqref{eq:langevin} does not properly describe the dynamics of the system.

Despite the practical difficulties in deriving a model of the 
form~\eqref{eq:langevin} from data, an approach based on the simultaneous 
evolution of $x_t$ and its discrete derivative $v_t$ seems still promising. In 
the following analysis we assume, as an ansatz, that the dynamics of $x_t$ can 
be approximated by a stationary stochastic process with discrete time, in which 
the stationary probability density function (pdf) of the process at time $t+1$ is entirely determined by the 
knowledge of the state $(x_{t},v_{t})$ of the system. From Eq.~\eqref{eq:velocity} it is clear 
that this amounts to say that
\begin{equation}
 p_n(x_{t+1} | \v{x}_{t}^{(n)})= p_2(x_{t+1} | \v{x}_{t}^{(2)}) \quad \quad \forall n\ge2\,;
\end{equation} 
we are thus assuming that the process is Markov, and that it ``keeps memory'' of 
the last two time-steps only. Let us notice that this assumption of a position-velocity
Markov process (p-vMP) may be consistent with a
two-dimensional structure of the dynamical attractor at the typical length-scales of the dynamics,
suggested by Fig.~\ref{fig2}(a) (we recall that $D\approx 1.3$ for ``large'' $L$).

With the above assumptions, we only need a practical procedure to determine 
$p(v_{t+1}|x_t,v_t)$ from data. A natural approach is the following:
\begin{enumerate}
 \item discretize the phase-space accessible to the vector $(x_t,v_t)$, by 
dividing it into a partition $\Gamma$ composed by $N_b^2$ rectangular cells, 
where $N_b$ is a suitable integer;
 \item consider an actual trajectory and look at the values assumed by $v_{t+1}$ 
at each time $t$ in which the state of the system falls into the $j$-th cell, 
$j=1,...,N_b^2$.
\end{enumerate}
  For each cell of the partition $\Gamma$ we can thus construct the histogram of 
the measured values of  $v_{t+1}$, conditioned to the state represented by that cell.
The range $(v_{min},v_{max})$ of the variable
$v$ is again divided into $N_b$ equal bins.

\begin{figure}
\includegraphics[width=0.99\linewidth]{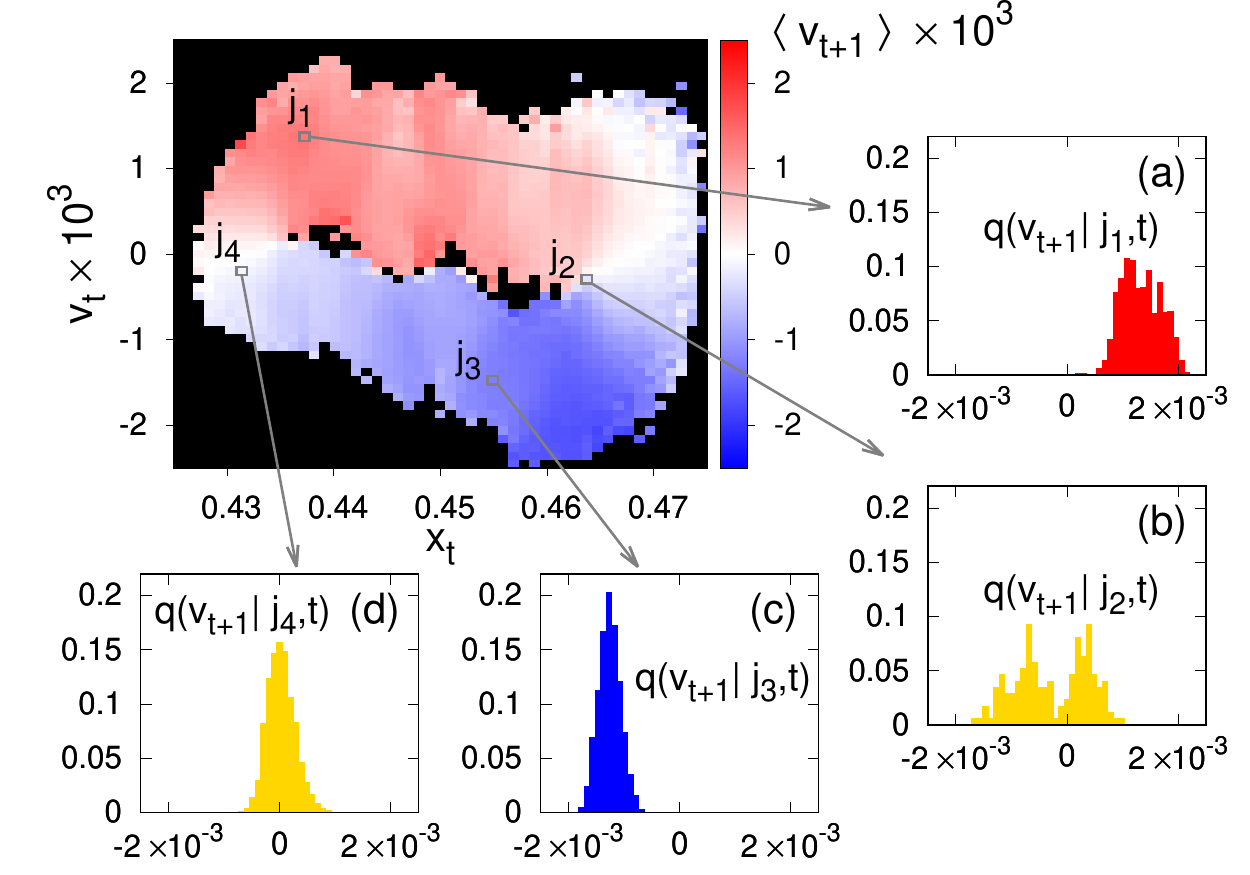}
 \caption{\label{fig3} Modeling a Markov dynamics. Main plot: for each cell of the (discretized) $(x_t,v_t)$ phase space, the average $\langle v_{t+1} \rangle$ is represented with the color code reported on the right side of the plot. Plots (a) - (d) show examples of the conditional empirical pdf $q(v_{t+1}|j,t)$, as discussed in the main text. Here $N=10^6$. The trajectory used to make this plot has a length $T=10^6$, the number of bins for the histograms and for the discretization of the phase space is $N_b=50$.}
\end{figure}

In Fig.~\ref{fig3} we give a graphical representation of our method. In the main 
plot, the (discretized) $(x_t,v_t)$ phase space is represented; the color of 
each cell is given by the average of $v_{t+1}$. Plots (a)-(d) show the 
histograms of $v_{t+1}$ for selected states, i.e the empirical pdfs 
$q(v_{t+1}|j,t)$. It is interesting to 
notice the qualitative difference between plot (b) and plot (d): in both cases 
$\langle v_{t+1} \rangle \simeq 0$, but in plot (d) the distribution has just 
one peak, centered in zero, and its spread is due to statistical fluctuations, 
whereas in plot (b) the histogram is clearly given by the superposition of two 
different states. Even if cases as that of plot (b) are quite rare, this is a 
first hint that our ``state identification'' by mean of $x_t$ and $v_t$ only is 
not exact. 

If the trajectory employed to build the
histograms is long enough, and the binning is sufficiently narrow, we can expect that
$q(v_{t+1} | j,t)$ is a fair approximation of the ``true'' stationary  conditional probability $p(v_{t+1}|x_t,v_t)$.
The rule:
\begin{enumerate}
 \item extract $v_{t+1}$ according to  $q(v_{t+1} | j,t)$;
 \item evolve $x_t$ by imposing $x_{t+1}=x_t+v_{t+1}$
\end{enumerate}
 defines a p-vMP on the (discretized) states of the system, which can be iterated to generate new
 trajectories mimicking the original dynamics. In the following section we will discuss how good
 this reconstruction is, and we will use it as a benchmark for the results of a machine-learning based approach.
 
 Let us notice that our method here is purely ``frequentistic'', meaning that we do not take advantage of any
 prior knowledge or assumption about the functional form of $q(v_{t+1} | j,t)$. If, by any chance, the reconstructed trajectory
 reaches a state which in the original trajectory has been explored very few times (less than a fixed threshold, 10 in our simulations),
 so that it is not possible to get a reliable prediction about $v_{t+1}$, the state of the system is re-initialized, according to the
 empirical distribution. However the frequency of this kind of events becomes negligible as soon as the length $T$ of the
 original trajectory is long enough (less than once in $10^4$ steps for $T\simeq 10^6$).

\subsection{\label{sec:machine} Machine learning (ML) approach}
In this section, we outline a machine learning approach to the construction of a 
stochastic model for the macroscopic dynamics. The key assumption is that the 
distribution of the variable $x(t)$ is a function of the past history of the 
variable itself: \begin{equation}
x_t\mbox{ drawn from }  p(\cdot|\{x_s\}_{s<t})
\end{equation}
just as one assumes when building the two-steps empirical p-vMP. Our purpose is 
to approximate $p$ with an empirical distribution $q$ given by a neural network. 
We remark that the role of the neural network is simply to provide a class of 
distributions (over the space of past histories $\{x_s\}_{s<t}$) which is large 
enough to fit any non pathological distribution $p$ in an efficient way (as 
insured by the universal approximation theorems)~\cite{mehlig2019artificial}. 
There are at least two relevant differences between the machine learning 
approach and the p-vMP. The first is that, unlike for the specific Markovian 
method presented before, the input values of the network-based recontruction of 
$p$ are (machine) continuous vectors. The second is that neural networks are 
naturally suited for processing long time series and we can easily condition 
forecasts $x_t$ to very long pieces of trajectories $\{x_s\}_{s<t}$. The reason 
is that the network is optimised to discard irrelevant information and, 
therefore, it may fit functions of high dimensional inputs with relatively low 
amount of data. Certain kind of neural networks can deal with indefinitely long 
input sequences in principle, but there is no reason to expect that arbitrarily 
long sequences of data can or need to be exploited. For this reason, in this 
paper we always base predictions on delay vectors with definite size 
$\v{x}^{(n)}_t=\{x_{t-n+1}\}_{s\leq t}$.

Likewise, even though we are working with time series, the most
suitable choice for our purpose is a feedforward neural network, as described in the following (see Appendix~\ref{app:ML}
for a minimal introduction to this kind of networks). Other tools, such as recurrent neural networks,
(e.g. LSTM, Long-Short Term Memory recurrent 
neural network~\cite{articleLSTM,Lei20}), can be very performing but would not
have allowed the efficient manipulation of the input vectors $\v{x}_t^{(n)}$ that we need to
investigate physical properties.


Our goal is to estimate
\begin{equation}   p_n(x_t|\{x_s\}_{t-n\leq 
s<t}) \end{equation}
from data. 
Given $p_n$, we define a Markov chain for $\v{x}^{(n)}$
\begin{equation}\begin{bmatrix} x_{t-1}\\ x_{t-2} \\ ...\\ x_{t-n}\end{bmatrix}\mapsto \begin{bmatrix}x^{forecast}_t\\ x_{t-1} \\ ...\\ x_{t-n+1}\end{bmatrix}\end{equation}
with 
\begin{equation} x^{forecast}_t \mbox{ drawn from } p_n(\cdot|\v{x}^{(n)}_{t-1})\end{equation}
with which we can approximate the true dynamics. In general, we expect that the reconstruction improves when increasing $n$. We can hope that, for some $n$, this procedure can provide a Markovian coarse grained model for the true dynamics. Notice that the $n=2$ case is conceptually equivalent to the p-vMP.

We have to approximate the distribution $p_n$ with a probability function 
defined on a finite number of elements with a feedforward neural network. Instead 
of binning the random variable $x$ itself, it is convenient to use the same 
approach as in the position-velocity Markov process by defining and attempt to approximate $p_n(v_t|\v{x}^{(n)}_{t-1})$.
This allows to achieve greater precision with fewer bins, since the typical scale of velocities is much smaller than the scale of positions, but it is not an otherwise obligated passage.


In order to achieve probabilistic predictions, the activation function of the neural network is chosen as a softmax, so that, overall, the neural network ``is'' the following probability function $Q_n$ (for delay $n$ and with an abuse of notation)
\begin{align}\label{Qn_}
\begin{split}
Q_n(j|\v{x}^{(n)}):&=\int_{v_{min}+(j-1)\,\Delta v}^{v_{min}+j\,\Delta v} dv\; q_n(v |\v{x}^{(n)})\\
&=\frac{1}{Z}\exp(H_j(\v{x}^{(n)}))\;\;\forall j=1,...,N_b+1
\end{split}
\end{align}
where, given $M$ layers, $H_a=\sum_{b=1}^{n_{M-1}} w^{(M)}_{ab} F^{(M-1)}_b+\theta^{(M)}_a$ are functions involving hidden layers of the neural network (see Appendix~\ref{app:ML} for details and notation). $\Delta v = (v_{max}-v_{min})/N_b$ is the bin size, where the extreme values $v_{max}$ and $v_{min}$ have to be computed empirically. $Q_n$ depends on a set of parameters $\Theta$ which can be fixed by minimizing a cost function with a gradient descent algorithm. A standard procedure for fitting the empirical distribution $Q_n$ is the minimization of the empirical cross entropy between the target distribution $P_n$ (which is the discretization of $p_n$) and $Q_n$, 
\begin{equation}\label{eq:empCost}
\hat C_n(\Omega_{train})=\sum_{i=1}^{T_{train}}\ln Q_n(I(v_{(i)})|\v{x}^{(n)}_{(i)}) 
\end{equation}
for a given training set $\Omega_{train}=\{\v{x}^{(n)}_{(i)},v_{(i)}\}_{i=1,...,T_{train}}$ where the operator $I$ assigns to every $v$ in the interval $(v_{min}, v_{max})$ the integer label
of the corresponding bin, ranging from 1 to $N_b$.
The expected value of $C$ is
\begin{equation}
\langle \hat C_n(\Omega)\rangle_{\Omega}=\int d\mu(\v{x}^{(n)})\sum_{j=1}^{N_b}\; P_n(j|\v{x}^{(n)})\;\ln Q_n(j|\v{x}^{(n)}) 
\end{equation}
This procedure is equivalent to minimizing the  quantity
\begin{equation}\label{disc_cost}C_n=\int d\mu(\v{x}^{(n)})\sum_{j=1}^{N_b}\,\Delta v\; \frac{P_n(j|\v{x}^{(n)})}{\Delta v}\;\ln \frac{Q_n(j|\v{x}^{(n)})}{\Delta v}
\end{equation}
since 
\begin{equation}
C_n=\hat C_n+\log(\Delta v).
\end{equation}
From now on, we will refer to~\eqref{disc_cost} when mentioning the cost function
 - unless otherwise specified - which will be always estimated empirically as in~\eqref{eq:empCost} on a test set $\Omega_{test}=\{\v{x}^{(n)}_{(i)},v_{(i)}\}_{i=1,...,T_{test}}$, statistically independent from $\Omega_{train}$.

For $\Delta v$ small, under physically reasonable assumptions, $C_n$ reduces to the relative entropy~\eqref{cost_cont} (see Section~\ref{sec:cost}), introduced as cost function for continuous-valued distribution, if we replace
\begin{equation}
\sum_{j=1}^{N_b}\Delta v \mapsto \int_{v_{min}}^{v_{max}} dv\mbox{ \quad and \quad } \frac{P_n}{\Delta v},\;\frac{Q_n}{\Delta v}\mapsto p_n,q_n.
\end{equation}

In order to be sure that the statistical approach is not a redundant machinery which can be replaced with a deterministic network, we can build a network for deterministic predictions and compare the outcomes of the two approaches. This network only differs from its stochastic counterpart in the last layers: it is a function $S_n$ (see~\ref{eqSn_app})
\begin{equation}\label{Sn_}
v^{forecast}=S_n(\v{x}^{(n)})
\end{equation}
which can be trained with a least square error procedure, i.e. minimizing
\begin{equation}\label{det_cost}
C_n(\Omega_{train})=\sum_{i=1}^{T_{train}}\|v_{(i)}-S_n(\v{x}^{(n)}_{(i)})\|^2
\end{equation}
with training set $\Omega_{train}=\{\v{x}^{(n)}_{(i)},v_{(i)}\}_{i=1,...,T_{train}}$.

All results in this article have been obtained with a feedforward neural network with two intermediate layers, each one with 300 hidden neurons. We also tried some different sizes but could not detect relevant differences in performance. The layers were fully connected and we employed the ReLU activation function everywhere least for the output layer. We trained the network with the ADAM~\cite{kingma2014adam} gradient descent algorithm.


\section{\label{sec:setting}Implementation details}

The methods introduced in the previous Section allow to build an effective 
dynamics from the analysis of trajectories. The outcomes of these protocols 
crucially depend on the setting of parameters such as the bin size and the 
length of the analyzed trajectory: in order to obtain meaningful results it is 
important to verify that our choices are sensible, and this task can be 
accomplished by a careful analysis of the cost function introduced in 
Section~\ref{sec:cost}. An important question is whether our results are 
descriptive of the system itself or whether they are only meaningful within the 
narrow context of our peculiar tool choice. With a careful analysis of the 
dependence of our methods on the setting parameters, we can definitely attribute 
results to the physics rather than the tool mechanics, with conventional 
limitations in confidence that are intrinsic in numerical studies. 

\begin{figure}
\includegraphics[width=0.99\linewidth]{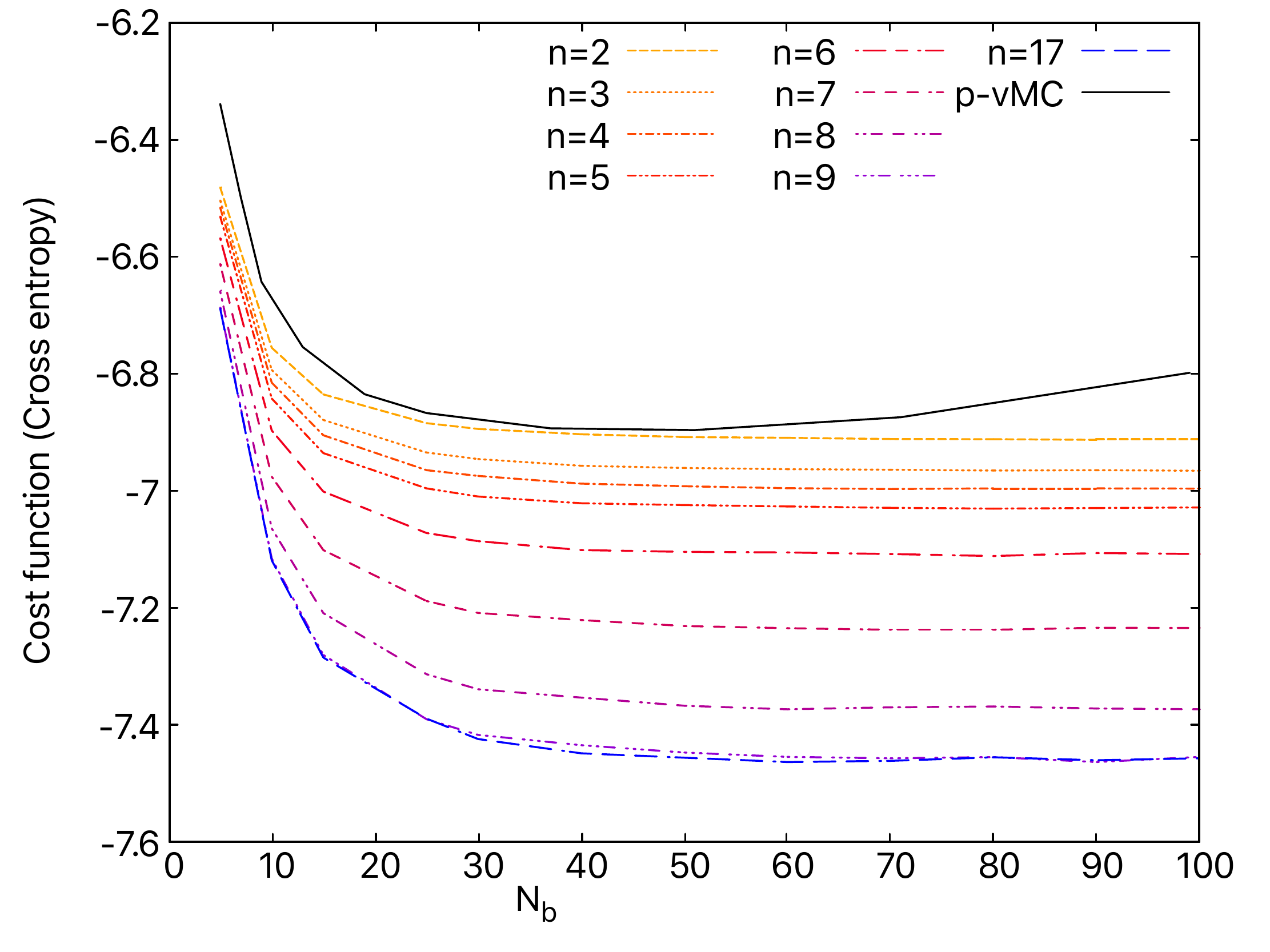}
 \caption{\label{fig:cost_and_bin} Dependence on the number of bins.
 Position velocity Markov process: black solid line shows the dependence of the cross entropy $C_{p\text{-}v}$ (Eq.~\eqref{eq:markovc}) on the number of bins $N_b$ in a p-vMP extracted from a trajectory of $T=10^6$ steps in a system of $N=10^6$ coupled maps. The optimal resolution is obtained for $N_b\simeq 50$. Neural network: the cost function (relative entropy $C_n$ ~\eqref{disc_cost}), computed on a test set of $T_{test}=5\cdot10^4$, reaches a plateau as the number of bins increases. The size of a bin is $\Delta v=(v_{max}-v_{min})/N_b=9.64\cdot10^{-5}/N_b$. The chart is has been obtained with $N=10^6$ maps; the training set size is $T_{train}=10^6$. Different delays $n$ are shown; notice the saturation for $n>9$.  }
\end{figure}

\subsection{The choice of binning}\label{sec:binning}

The choice of the number of bins $N_b$ corresponding to each dimension is not just 
a technical point, but is related to the physics of the system. Indeed,
$\Delta v=(v_{max}-v_{min})/N_b$ 
fixes the lowest scale at which increments $x_{t+1}-x_t$ can be resolved or, 
equivalently, the order of magnitude of the typical one-step forecasting error: 
all phenomenology taking place below this scale is implicitly discarded by the 
model. This is particularly relevant when modelling systems with multiple 
spatial scales, such as the one we are examining~\cite{cencini99}.

In the  p-vMP approach the number $N_b$ of bins per linear dimension plays a 
main role, and it has to be chosen with particular care. On the one hand, this 
parameter determines  our ability to ``resolve'' the state of the system, i.e. 
the point of the position-velocity phase-space in which the system is found; 
indeed, with our protocol such phase-space is divided into rectangular cells 
whose size scales as $N_b^{-2}$. On the other hand, $N_b$ is also the number of 
bins used for the discretization of the conditional pdf $p(v_{t+1}|j,t)$, and 
therefore it determines the precision of our forecasting.

If one had access to arbitrarily long trajectories (and arbitrary machine 
precision), the quality of the reconstruction would be directly determined by 
$N_b$: the larger the number of bins, the better our ability to describe the 
system. Since in practice one always deal with finite trajectories, choosing too 
large values of $N_b$ is, conversely, detrimental: indeed for our method to work 
properly it is crucial that most cells along the attractor are visited by the 
``original'' trajectory many times, enough to build a reliable histogram for the 
conditional pdf  $p(v_{t+1}|j,t)$; if the binning is too small, this condition 
will fail to be fulfilled and the quality of our prediction will result quite 
poor, despite the high level of resolution.

The above reasoning can be verified quantitatively by looking at a cross entropy
quite similar to that defined by Eq.~\eqref{cost_cont} and discussed in
Section~\ref{sec:cost}, i.e.
\begin{equation}
\label{eq:markovc}
C_{p\text{-}v}=-\int  d\mu(x_t,v_t)\int dv_{t+1}\, p(v_{t+1}|x_t,v_t)\;\ln q(v_{t+1}|x_t,v_t)\,,
\end{equation} 
which is equivalent, by construction, to $C_2$. Fig.~\ref{fig:cost_and_bin} 
(black solid curve) shows the value of $C_{p\text{-}v}$ as a function of $N_b$, when a 
trajectory of $T=10^6$ time steps is considered. We observe that the optimal 
number of bins is around 50, since after that  value the cross entropy starts 
increasing, a clear hint that the quality of the reconstruction is reducing.

As anticipated, the network-based model requires a binning on the velocities, just like 
the p-vMP approach. In order to analyse the role of $\Delta v$, assume our training set is large 
enough (e.g. $T_{train}\approx10^6$) to avoid significant dependence on its 
size - the dependence on $T_{train}$ will be discussed later. Then, for any fixed $n$, we can 
compute the cost function $C_n$ as a function on $\log(\Delta v)$. We can see 
from Fig.~\ref{fig:cost_and_bin} that if $\Delta v$ is small enough (i.e. 
$N_b$ is sufficiently large), $C_n$ reaches convergence. This means that the sum 
appearing in the cost function (see equation ~\eqref{disc_cost}) is converging - in a 
subset of the attractor of $\v{x}^{(n)}$ whose probability is close to 1 - to a 
well-defined integral in Riemann-sense. 

Let us notice that in this case there is no upper limit on the choice of $N_b$, 
at variance with the p-vMP approach, since now $N_b$ does not play any role in 
the identification of the state. This explains the difference between the curves 
representing  $C_2$ and $C_{p\text{-}v}$ in Fig.~\ref{fig:cost_and_bin}: they are 
qualitatively similar for $N_b<50$; then the former reaches a plateau, the 
latter starts increasing for the reasons discussed above. Let us notice that the 
optimal values reached by these two curves are very similar: this is not 
surprising, since conditioning on of $x_t$ and $x_{t-1}$, as it is done in the 
ML approach with delay $n=2$, should be equivalent to conditioning on $x_t$ and 
$v_t=x_t-x_{t-1}$.

Figure~\ref{fig:cost_and_bin} is the first confirmation that our stochastic approach is more appropriate than a deterministic one. Indeed, it shows that the fraction of bins with non-zero probability (for any given $\v{x}^{(n)}$) does not decreases with $N_b$. The latter case, which would correspond to $C_n$ being a decreasing function of $\Delta v$, would have been observed for a deterministic-looking dynamics (or a distribution with several very thin peaks) where, for any $\v{x}^{(n)}$, there should be single ``active'' bin with non null probability, provided that $n$ is large enough that we can write $x_t=F(\v{x}^{(n)}_t)$ with some deterministic function $F$. However it, it is important to consider that the plateau (of $C$ vs $N_b$) is not to be interpreted as evidence that the observed dynamics is truly stochastic, the smoothness of $q_n$ may be a result of $n$ being too short (or not having enough data or our machine learning procedure not being good enough): we know by construction that this is our scenario. If we could use a delay $n$ comparable to $N$ and both an implausible large neural networks and a prohibitive long training length, we could have unveiled the decreasing behaviour $C_n$ (with respect to $\Delta v$) associated to the deterministic nature of the coupled maps system. One might also notice that, in Fig.~\ref{fig:cost_and_bin}, curves with $n=9$ and $n=17$ overlap, implying the existence of a second plateau in $C$ vs $n$: this will be throughly discussed later. 

The ML analysis shown in the following Sections has been done by adopting $N_b=50$, for the sake of consistency with the p-vMC. Further refinement, say $N_b=100$, does not bear much greater computational hardness. Our analysis suggests that this binning was sufficient for good model building.


As a broader methodological note, it is important to highlight that, in the general scenario of modelling a $D$-dimensional dynamical variable, the number of bins would scale as $N_b^D$ ($N_b$=bin number per linear dimension), which is not out reach for up to date machine learning technology, as long as $D$ is reasonable low and one has enough data, but could otherwise result in the curse of dimensionality.



\begin{figure}
\includegraphics[width=0.99\linewidth]{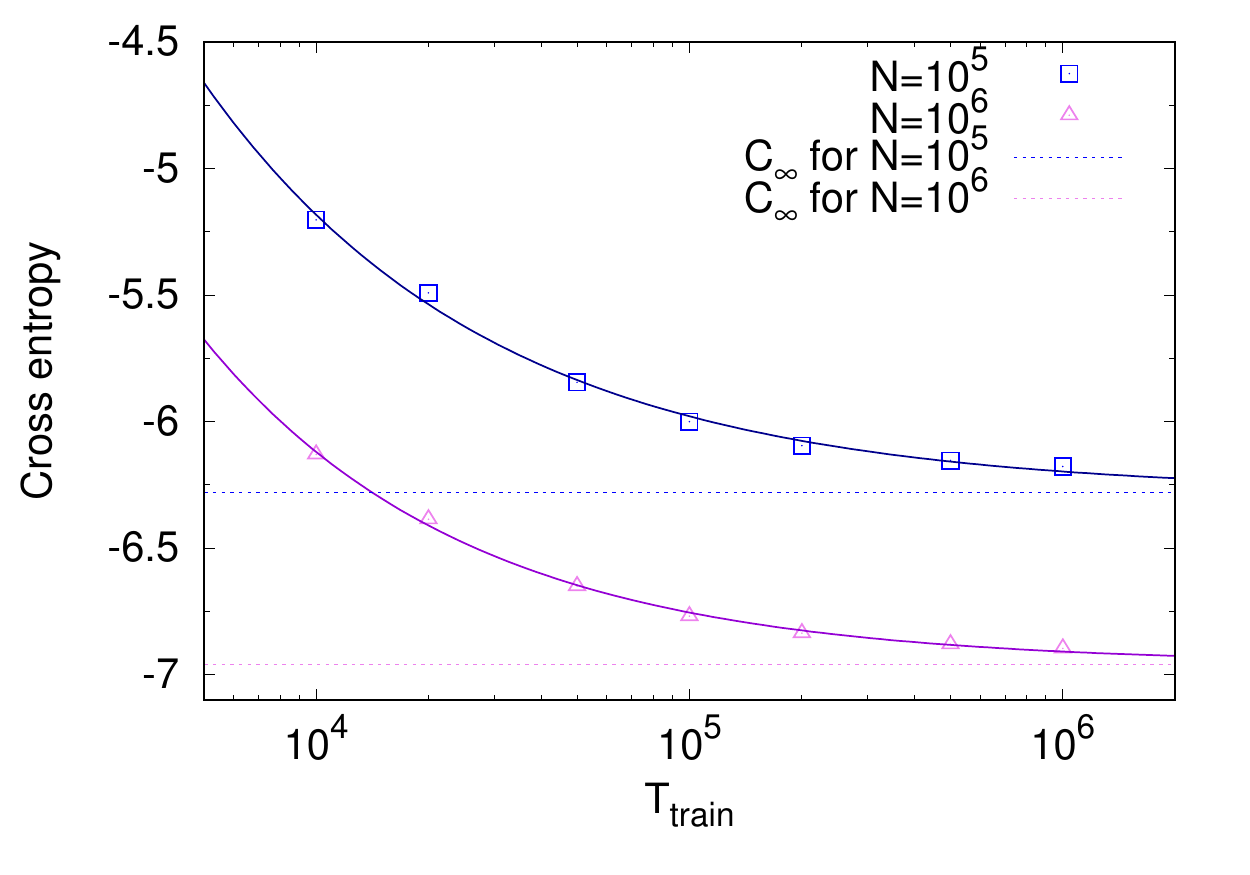}
 \caption{\label{fig:markov_t}Dependence on $T_{train}$ in the p-vMP analysis.
 For two different choices of $N$, the behaviour of the cost function~\eqref{eq:markovc} as a function
 of the length of the analysed trajectory is shown. Solid curves are obtained with a fit of the functional form~\eqref{eq:cpower}, whereas dashed lines represent the inferred value of $C_{\infty}$. Here $N_b=50$.  From the power-law fit we find $C_\infty=-6.28$, $\alpha=-0.562$ for 
$N=10^5$ and $C_\infty=-6.96$, $\alpha=-0.615$ for $N=10^6$. }
\end{figure}

\begin{figure}
\includegraphics[width=0.99\linewidth]{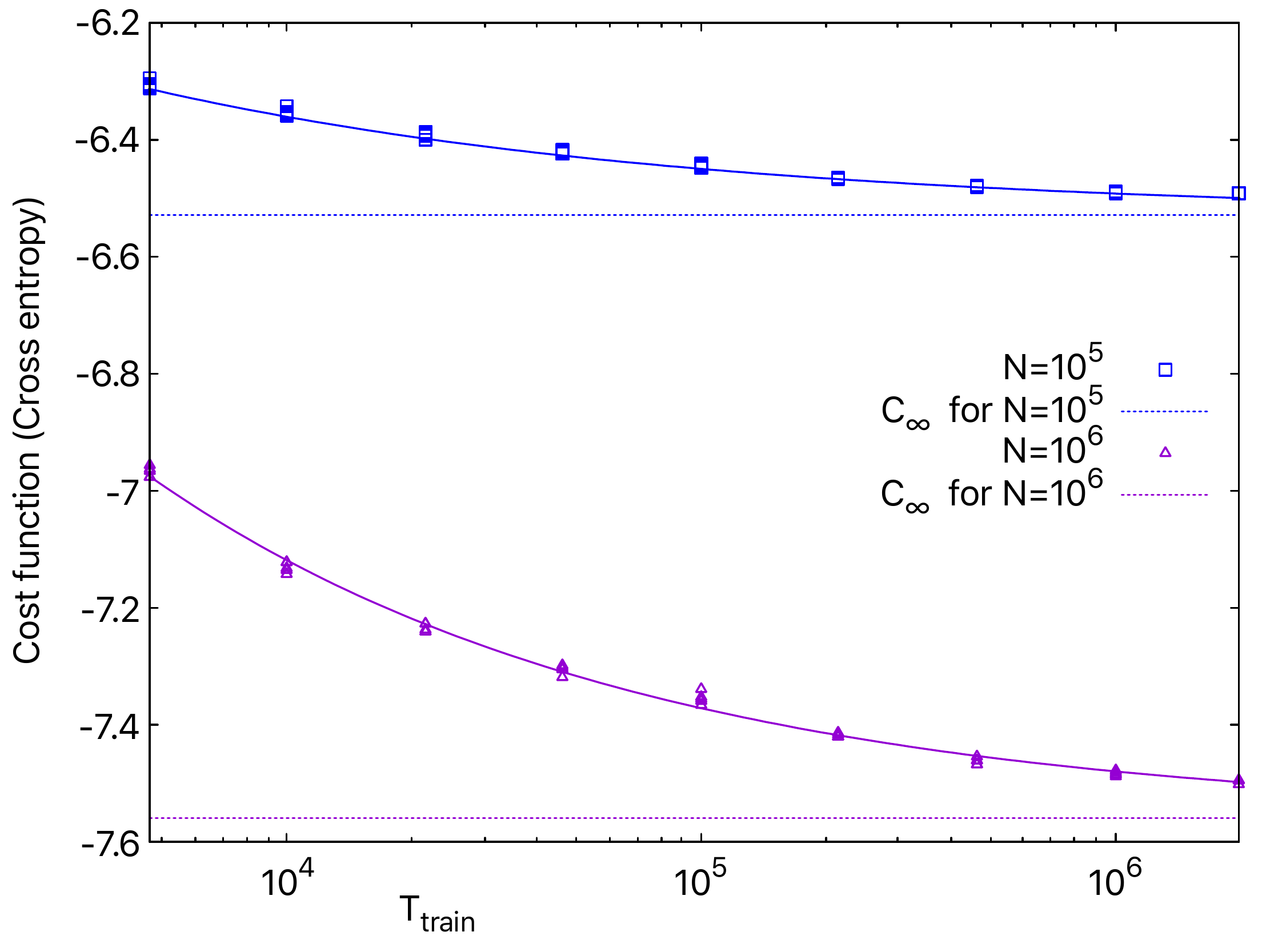}
 \caption{\label{fig:fit_t_train} Dependence on $T_{train}$ in the ML analysis. 
The cost function decreases in $T_{train}$ as a power law. We have grouped 
values for different $n>\tilde n$ both in $N=10^5$ and $N=10^6$ cases. Dashed 
lines show extrapolated asymptotic values of the cost function $C_{\infty}$. We 
have used $T_{test}=10^6$ and $N_b=50$. From the power-law fit we find $C_\infty=-6.53$, $\alpha=-0.325$ for 
$N=10^5$ and $C_\infty=-7.59$, $\alpha=-0.368$ for $N=10^6$.
} \end{figure}

\subsection{Dependence on the training trajectory and asymptotic extrapolations}

The approaches presented in Section~\ref{sec:dynamics} are based on the extrapolation of relevant information from data, in order to make reliable predictions. The quality of the results clearly depends on the amount of available data, i.e. on the length $T_{train}$ of the original trajectory employed to infer the conditional probabilities of the p-vMP and to optimize the internal weights of the neural network during the ``training'' phase. The value of the cost function defined by Eq.~\eqref{cost_cont} is thus expected to decay with $T_{train}$. We observe that in all considered cases such decay is well described by a power law
\begin{equation}\label{eq:cpower}
C_n(N,N_b)\approx C_{\infty}(n,N,N_b)+\frac{c(n,N,N_b)}{T_{train}^{\alpha(n,N,N_b)}}
\end{equation}
where $C_{\infty}$, $c$ and $\alpha>0$ are parameters which will depend, in general, on $N_b$, $N$ and $n$, but we will drop dependencies for simplicity (as before, the cost function defined by Eq.~\eqref{eq:markovc} can be seen as equivalent to $C_2$).

Figure~\eqref{fig:markov_t} shows the case of the p-vMP, for two choices of $N$. Assuming that Eq.~\eqref{eq:cpower} holds asymptotically, we expect that no significant improvement in the reconstruction would be achieved by considering values of $T_{train}$ larger than $10^6$ (at least with this choice of the binning). Consistently, in the following we will always keep $T_{train}=10^6$.

\begin{figure}
\includegraphics[width=0.99\linewidth]{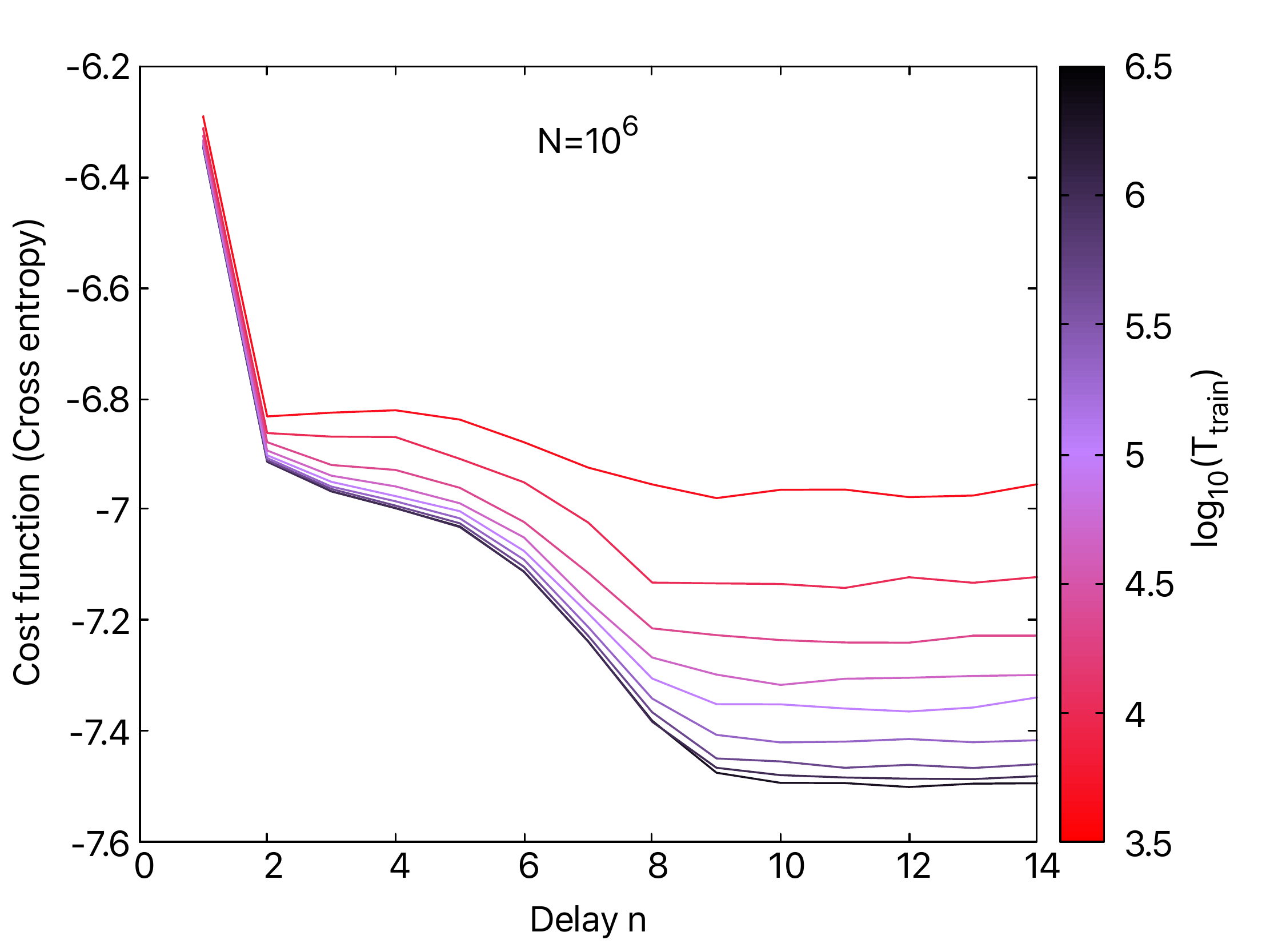}
 \caption{\label{fig:t_train} 
For $N=10^6$ maps, we show as the value of $\tilde n\approx 10$ emerges as $T_{train}$ increases . Here $T_{test}=10^6$. }
\end{figure}

In order to control for the effect of the training size in the ML approach we can plot the cost function as a function of $T_{train}$ for different values of $N$  (see Figg.~\ref{fig:fit_t_train} and ~\ref{fig:t_train} for more details on $N=10^6$ case).  While $\alpha(n,N,N_b)$ and $c(n,N,N_b)$ are likely to be strongly dependent on the machine learning algorithm, the extrapolated $T_{train}\to\infty$ asymptotic value $C_{\infty}(n,N,N_b)$ should be an estimator for a procedure-independent observable, with a (hopefully small) bias deriving from the specific machine learning protocol: we know that, for fixed $N$, $C_n(N,N_b)$ reaches a plateau when 
$N_b$ is large and $n>\tilde{n}$ and we can interpret this as evidence of an underlying effective low dimensional macroscopic Markovian structure in the original system. This Markovian structure is identified by a transition probability distribution which is well approximated by the empirical function $q_n$ as defined in~\eqref{disc_cost}. If this is true, then the quantity (see Section~\ref{sec:cost}):
\begin{equation}C_{\infty}\approx -\int d\mu(\v{x}^{(n)})\int dx\, p_n(x|\v{x}^{(n)})\;\ln q_n(x|\v{x}^{(n)})\mbox{ for }N\gg n>\tilde n\end{equation}
is a physical observable describing the cost function of the coarse grained process with respect to the true dynamics. 

Note that the $T_{train}\to \infty$ limit we have mentioned is physically misleading. Indeed, we have  no reason to believe that our results will hold for arbitrary large neural network (for instance a network with $\gg N$ many neurons) and in the actual $T_{train}\to \infty$ limit. Indeed, it is argued~\cite{cencini2000chaos,falcioni2005properties} that the apparent nature of a systems depends of the length of observations. In particular, one may not distinguish a periodic dynamics with very large period, genuine chaos or a stochastic dynamics without enough data; pseudo-convergences of the Shannon entropy rate can hide the true nature of the system. This is our case, since we derived a stochastic system from equations which are deterministic by construction. Nevertheless, this is not a problem, since we are not aiming at reconstructing the ``true'' system but rather to provide a coarse grained characterization. If we are confident that the reconstruction is efficient, then we can say that $p\approx q$, so that, if $N\gg n\geq\tilde n$
\begin{equation}
C_{\infty}\approx h= \int d\mu(\v{x}^{(n)})\int dx\, p_n(x|\v{x}^{(n)})\;\ln p_n(x|\v{x}^{(n)})
\end{equation}
which represents the Shannon entropy rate of the coarse grained dynamics in the Markovian approximation and quantifies how much information we are losing, in terms of one-step forecasts, by coarse graining. Therefore, we stress that this machine learning method, by providing an explicit - albeit approximated - expression for $p_n$, can overcome standard difficulties in quantifying the Shannon entropy rate of the coarse grained process.

An empirical confirmation supporting the physical interpretation of machine learning results will be given in spectrum analysis section.

\section{\label{sec:discussion} Results and discussion}

The ML approach introduced in Section~\ref{sec:machine} aims at mimicking the dynamics produced by Eq.~\eqref{eq:xt} by the analysis of time series of data, without any additional information about the generating model. In this Section we show the outcomes of this approach, and we compare them to the results of the purely frequentistic p-vMP method. In this way, on the one hand, it will be possible to quantify the predictive power of our ML approach with respect to a ``benchmark'' strategy, whose working principles are completely clear; on the other hand, as we will discuss in the following, the comparison of the two methods helps understanding which additional features of the generating dynamics are caught by the ML, and disregarded by more straightforward approaches.
As anticipated in Section~\ref{sec:cost}, the main tools for our validation tests are the cross entropy and the analysis of Fourier frequency spectra of trajectories generated by the considered methods.

\subsection{Cost function vs delay and model building}

\begin{figure}
\includegraphics[width=0.99\linewidth]{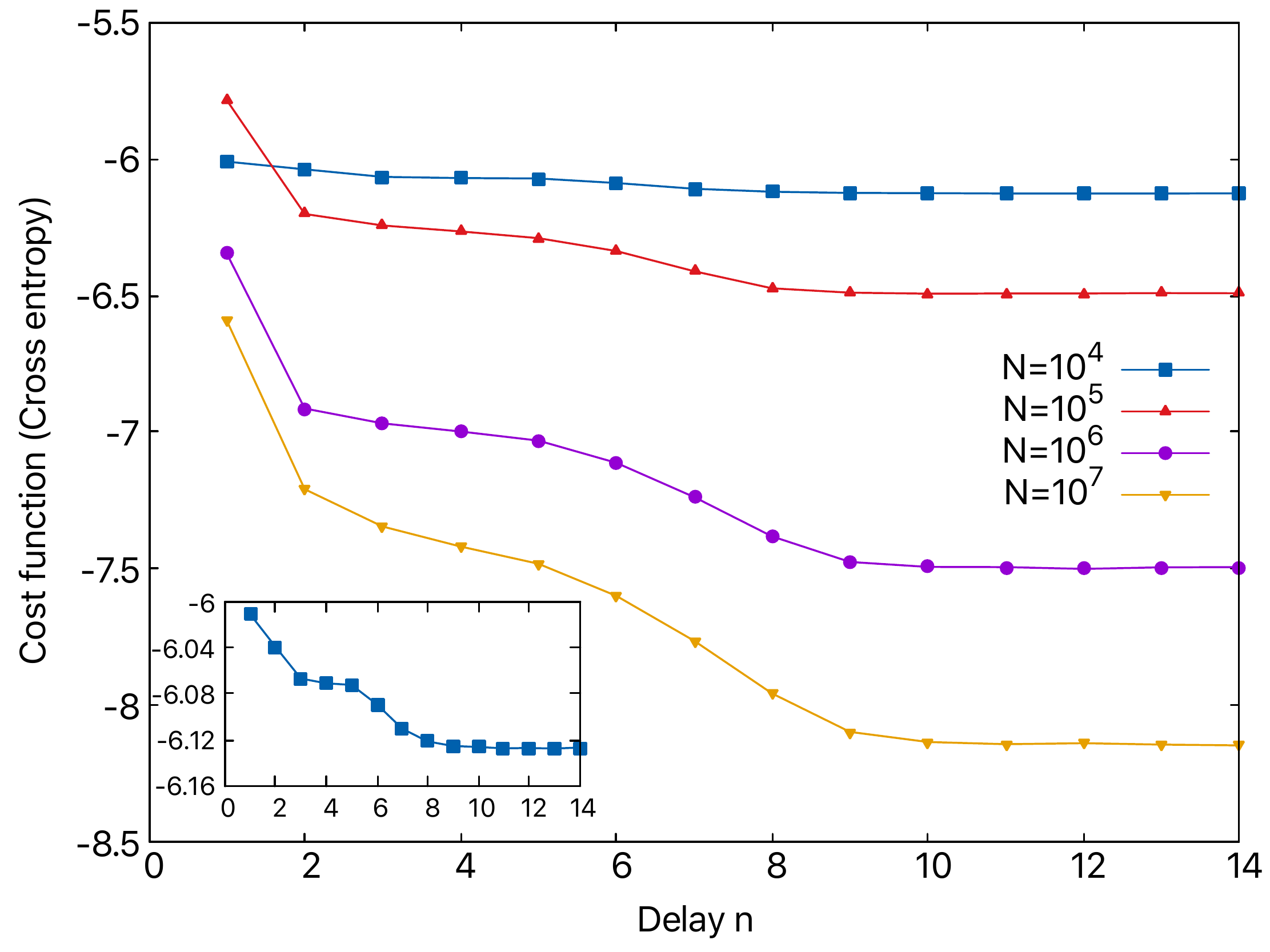}
 \caption{\label{fig:cost_and_delay} 
The cost function (relative entropy $C_n$~\eqref{disc_cost}), computed on a test set of $10^6$, decreases with $n$ for all $N$. After $\tilde n\approx 10$, a plateau is visible for any value of $N$. The training set size is $T_{train}=2\cdot10^6$ for $N=10^4,\;10^5,\,N^6$ and $10^6$ for $N=10^7$. The inset shows a magnification of $N=10^4$ curve.  }
\end{figure}

In the ML approach, the dependence of the test cost function on the delay $n$ is expected to be, in principle, the result of two effects pointing in opposite directions. As $n$ increases, the dimension of the input increases and so does the amount of training data needed: this effect alone would make the cost function an increasing function of $n$. Conversely, as $n$ increases, more information can be extracted from past history of the dynamic variable and, hence, this would make the cost function a non-increasing function of $n$. In practice, as shown in Fig.~\ref{fig:cost_and_delay}, with training length chosen $T_{train}=10^6$, the first effect is barely noticeable and we can focus on the second. For small values of $n$, the function $C_n$ decreases; after some value $\tilde n$ it reaches a plateau. Remarkably, the value of $\tilde n$, which is found to be  $\simeq 8-10$, is very similar for all different number of maps we have examined, i.e. $N=10^4,\;10^5,\;10^6,\;10^7$.

If we exclude strong effects by either finite training length or network size, the plateau after $\tilde n(N)$ implies that no more information can be extracted by increasing the delay, given our resolution. This implies that
\begin{equation}
P_n(x_t|\v{x}^{(n)}_{t-1})\approx P_m(x_t|\v{x}^{(m)}_{t-1})\;\mbox{ if }m,n>\tilde n
\end{equation}
i.e. the system appears to be well described by a Markov process at our level of coarse graining. Hence, for any $n\geq\tilde n$, we define the Markovian machine learning model as the Markovian dynamics given by transition matrix $Q_n$ as defined in Section~\ref{sec:machine}.  This is, by construction, the best coarse grained model, in terms of one-step likelihood, we could obtain with our procedure. We will show later that this model is efficient in reproducing the dynamics on all timescales and for any $N$ we have considered.

In order to obtain all figures in this section, we have trained a single neural network for each $n$ and $N$. The length of the training sequence is $T_{train}=2\cdot10^6$ for $N=10^4,\;10^5,\,10^6$ and and $T_{train}=10^6$ for $N=10^7$. For each pair $n,\;N$, we have kept the best performing set of neural-network parameters found in 100 epochs (the number of epoch is, informally, the number of times the network is optimized over the same training set. See Appendix~\ref{app:ML}). The performance has been evaluated with a test set of size $T_{test}=10^6$ for all $N$.

\begin{figure}
\includegraphics[width=0.99\linewidth]{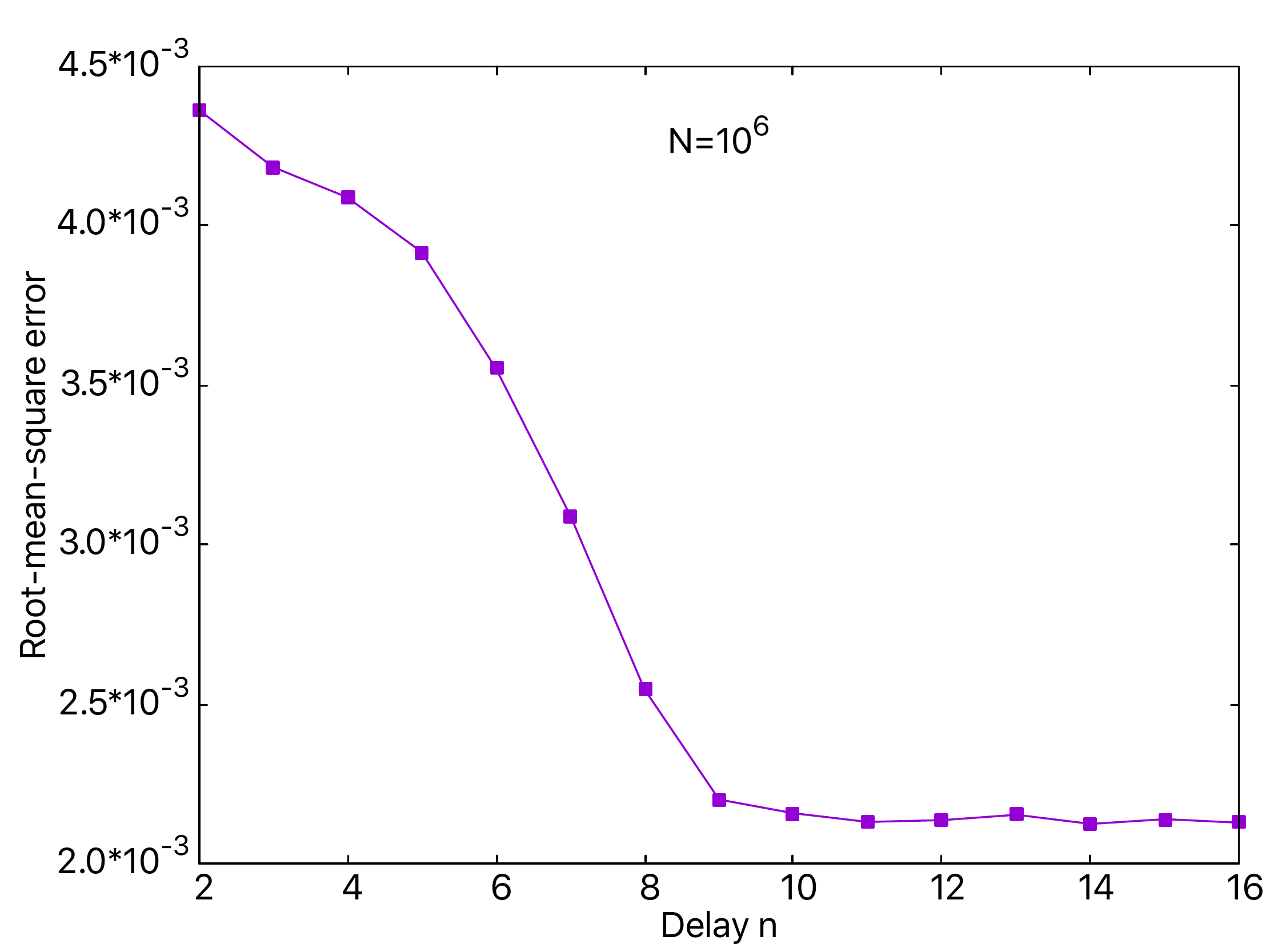}
 \caption{\label{fig:cost_and_delay_det} 
  Deterministic neural network model. Square root of the test cost function (mean square error $C_n^{det}$ in equation~\eqref{det_cost}) decreases with $n$, shown for $N=10^6$. After $\tilde n\approx 10$, a plateau is visible. Test set size $T_{test}=50000$ while train set size $T_{train}=2\cdot10^6$. Compare plateau height with scale crossover in Fig.~\ref{fig2}. }
\end{figure}

In order to further confirm that the delay $\tilde n$ pertains to the dynamics and not a machine-learning construct, in Fig.~\ref{fig:cost_and_delay_det} we plot the (square root of) mean-square-error cost function ($N=10^6$), calculated with the deterministic network described above: the behaviour is the same and the plateau begins around the same values of $n$. Note that the plateau height is $\approx 10^{-3}$, which is the scale crossover to high dimensional dynamics as highlighted by Grassberger-Procaccia analysis (see Fig.~\ref{fig2}).

\begin{figure*}
\includegraphics[width=0.99\linewidth]{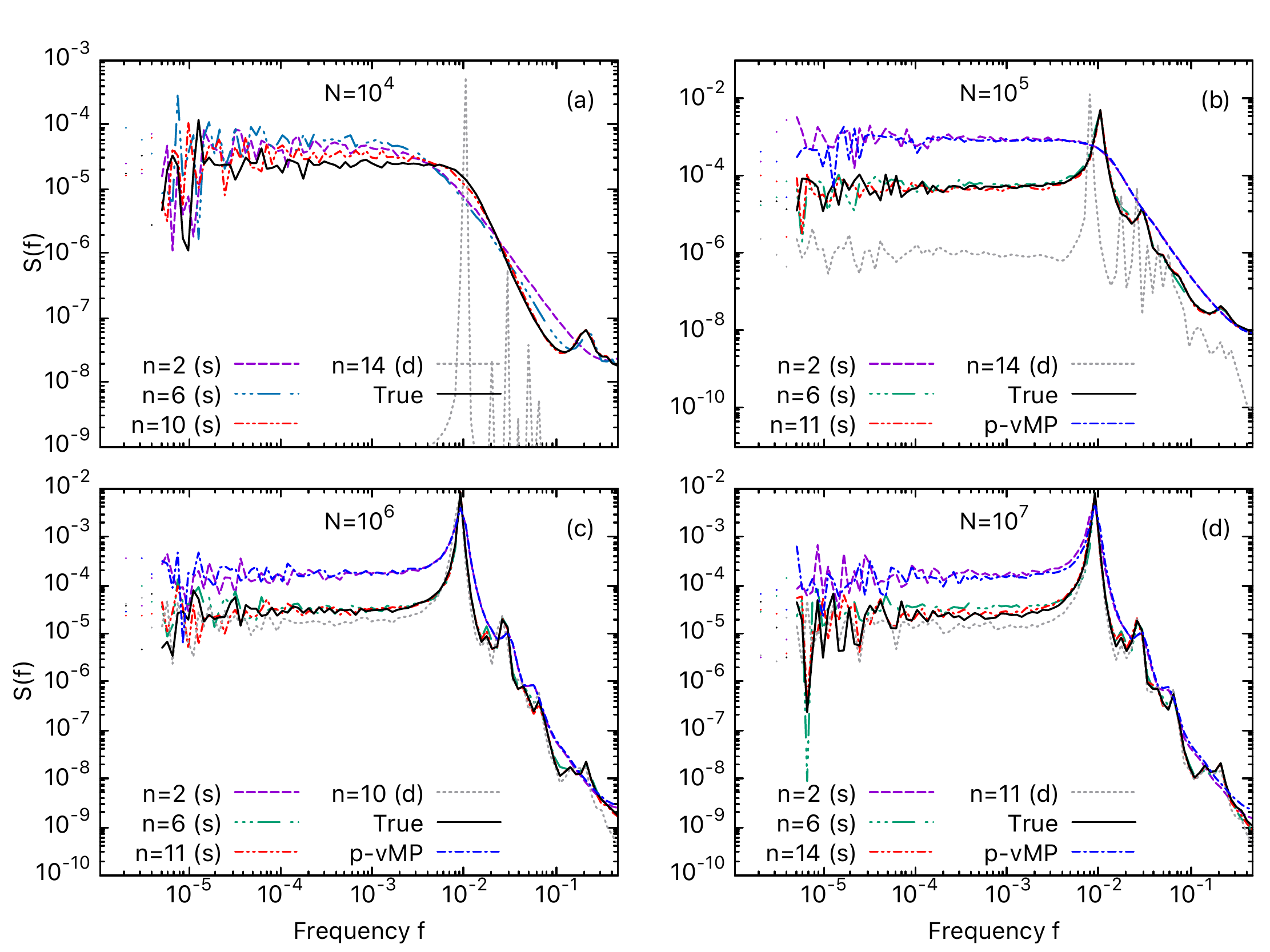}
 \caption{\label{fig:spettri} Fourier spectra for different number of maps $N$. In each panel, the true spectrum is compared to its counterparts from ML and p-vMP (but for panel (a), in whcih the p-vMP method is too unstable to allow for a trajectory reconstruction). Stochastic model (s): the frequency spectrum of the delay $n=2$ neural network model almost perfectly overlaps with the position-velocity Markov process. They both miss the main frequency peak for $N=10^5$, even if they are able to catch the main features of the dynamics for larger values of $N$ (see Section~\ref{sec:subdiscussion} for a discussion on this point). With delay $n \gtrsim 10$, the reconstruction is always good. Deterministic model (d): failed reconstruction, to some degree, can be seen at all scales even with $n\geq \tilde n$, implying that the stochasticity is a key feature; on the other hand, the deterministic model captures most frequency peaks. Moreover, it greatly improves with $N$ and around $N=10^6$. For all charts $T_{train}=T_{test}=10^6$.} 
\end{figure*}


\subsection{Spectral analysis}

For evaluating long term reconstruction, we have compared the Fourier spectra of the original system against that of the proposed models; the results are shown in Fig.~\ref{fig:spettri}. For the ML approach with $n=2$ or $n=3$ and the position-velocity Markov process, we get a fair agreement at high frequency $f$, and a correct detection of the peak at $f\simeq 10^{-2}$, if the number of maps  is sufficiently large ($N \gtrsim 10^6$). For $N=10^5$ these methods are not able to catch the oscillatory behaviour of the dynamics, as can be understood by the absence of the above mentioned peak in the dynamics. A proper description of the low-frequency dynamics is never accomplished.

As for the stochastic neural-network based model, we can see that for $n\geq \tilde n$, good reconstruction has been achieved at all frequencies (see Fig.~\ref{fig:spettri}). Notably, a delay of $n=6$ seems to be enough - if $N\geq 10^6$ - for the correct modelling of the low frequencies. For shorter delays, there seems to be not enough information and the model overestimates stochasticity, generating spurious long oscillations which would not be allowed by the coupled maps dynamics.
The deterministic models, on the other hand, always underperform their stochastic counterparts: most frequency peaks are captured, 
 while low frequency oscillations are damped. 
Reasonably, a deterministic description efficiently captures quasi-periodic behaviours but fails to account for high dimensional residual dynamics, which - in an effective description - acts as noise by widening  peaks and generating rich low frequency dynamics. Case $N=10^4$ is instructive: the original dynamics lacks any quasi-periodic behaviour and the model critically fails by collapsing unresolved noise into a single peak. Still, as $N$ increases, the deterministic description greatly improves and, at $N=10^6$, the performance good enough provide an alternative option to its stochastic counterpart, at least at large scales. 

As a technical remark, in identical settings ($n$, $N$, $T_{train}$ etc.), even the same short term performances do not always imply comparable long term identical results (we remind that the result of the training process is not deterministic). The variability is low for the stochastic nets but noticeable for deterministic one. Since we were only interested in showing that the problem can be solved but not in machine learning algorithms per se, we only displayed performing cases in Fig.~\ref{fig:spettri}. This is a very well know issue and the underlying reason is probably the choice of hyper-parameters: a more efficient machine learning technique can most likely solve the issue, but it is beyond the scope of this work.

\begin{figure}
\includegraphics[width=0.99\linewidth]{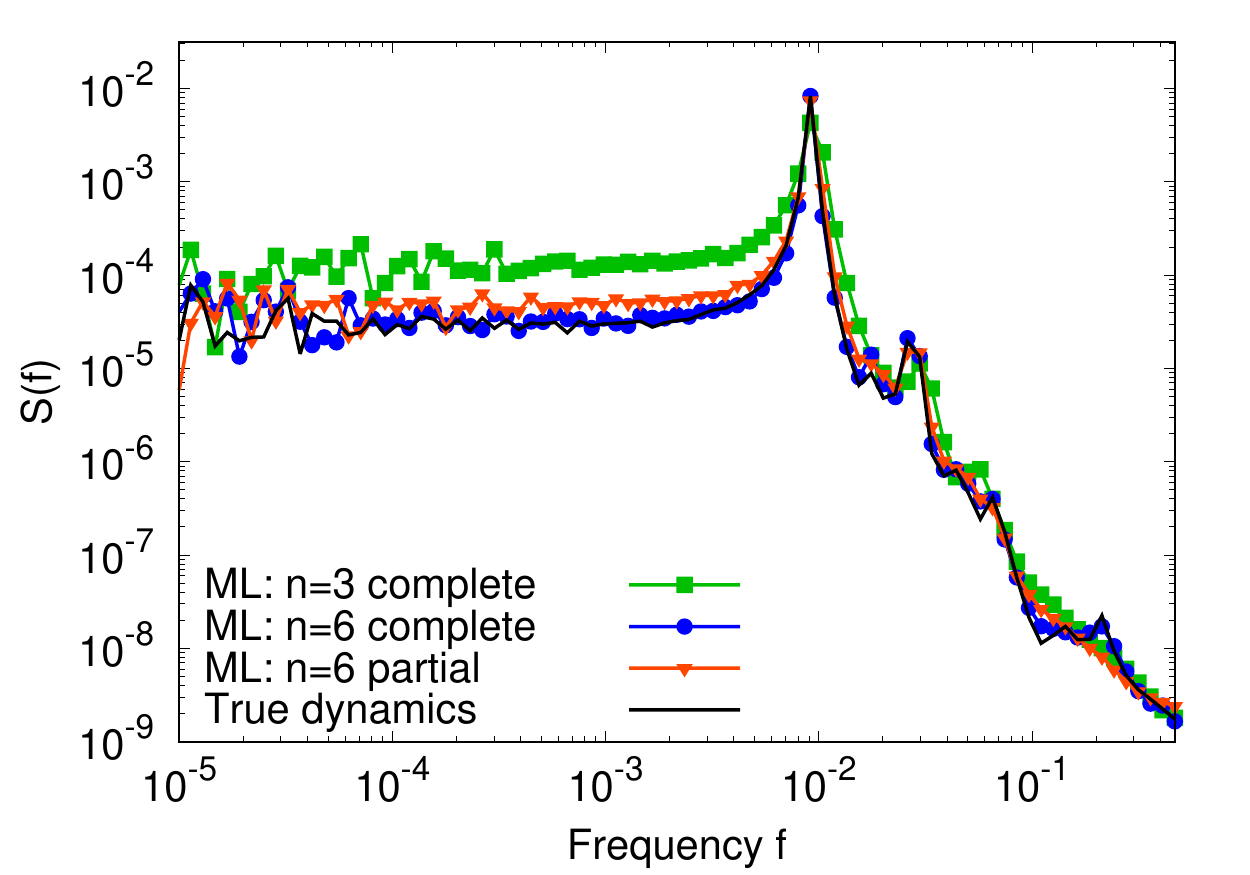}
 \caption{\label{fig:spectrum3} Proper choice of the variables. The spectrum of the actual dynamics (black solid line) is compared to those obtained with a ML approach with delay  $n=3$ (green squares) and $n=6$ (red circles). Blue triangles show the result of a reconstruction obtained by considering only the ``partial'' embedding vector $(x_{t-1},x_{t-2},x_{t-6})$. The good agreement of the last curve with the original dynamics shows that the entire knowledge of the $n=6$ delay vector is redundant. Parameters of the reconstruction: $n=10$, $T_{train}=10^6$, $T_{test}=10^6$, $N=10^6$.}
\end{figure}

\subsection{Proper selection of the variables}
\label{sec:tre_n}

The results discussed above clearly show that the stochastic ML approach is able to catch the essential low-frequency features of the original trajectories when $n\gtrsim6$, a threshold which is quite independent of $N$. Apparently, this fact seems to suggest that the true dynamics needs, at least, a 6-dimensional description to be correctly reproduced. On the other hand, there is actually no valid reason to believe that \textit{all} elements of the embedding vector are equally important to the reconstruction of the dynamics, and it is even possible, in principle, that the knowledge of some of them is redundant.
Some hint on this point may be gained by a careful analysis of Fig.~\ref{fig:cost_and_delay}. The decrease rate of the cross entropy is not constant along the curve: the inclusion of some specific elements seems to be particularly relevant. For instance, passing from $n=5$ to $n=6$ improves the reconstruction ``more'' than passing from $n=3$ to $n=4$ (if we use the cross entropy as a metric for the quality of the reconstruction).

It is natural to wonder what happens if one does not consider the \textit{entire} embedding vector, i.e. if only some elements are passed to the neural network. Fig.~\ref{fig:spectrum3} shows the spectrum that is obtained if the vector $(x_{t-1},x_{t-2},x_{t-6})$ is used to infer the probability distribution of $x_t$: its agreement with the original dynamics is almost as good as that obtained with the complete $n=6$ embedding vector. This fact may imply that the effective dimension of the attractor of the coarse-grained dynamics is actually close to 3 in the case under study.

\begin{figure}
\includegraphics[width=0.99\linewidth]{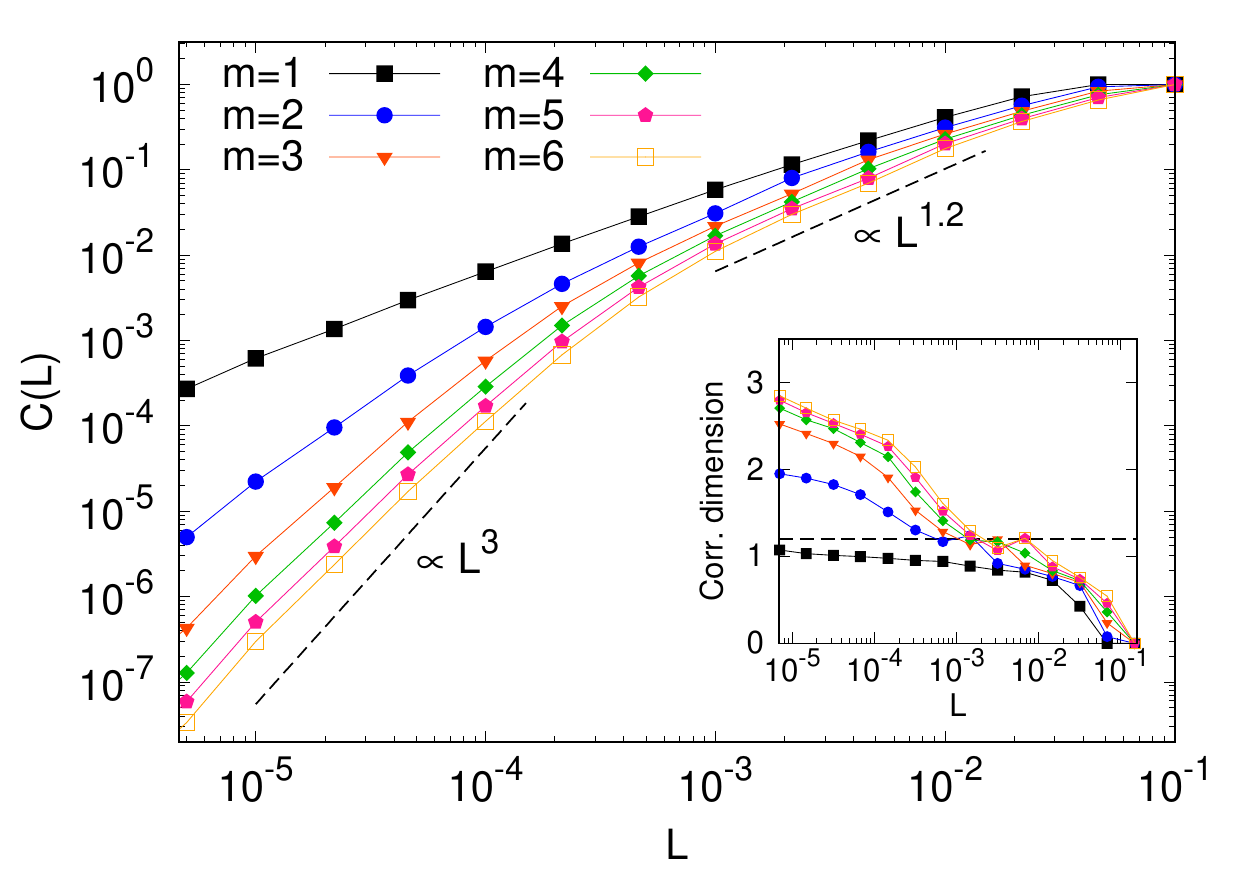}
 \caption{\label{fig:GPdet} GP analysis of a (deterministic) ML-generated trajectory. As in Fig.~\ref{fig2}(a), in the main plot we show the correlation~\eqref{eq:gp} as a function of the lengthscale $L$, for different values of the embedding dimension; the inset reports the corresponding values of the attractor dimension. At least two different regimes can be individuated: a small-scale behaviour with dimension between 2 and 3 and a high-scale one, similar to that shown in Fig.~\ref{fig2}. Parameters of the reconstruction: $n=10$, $T_{train}=10^6$, $T_{test}=8\cdot 10^6$, $N=10^6$.}
\end{figure}
\subsection{Discussion}
\label{sec:subdiscussion}

By solely looking at Fig.~\ref{fig2}, reporting the Grassberger-Procaccia analysis of the original trajectory, one may expect that a coarse-grained dynamics with dimension $\simeq 1.3$ would suffice to catch the macroscopic behaviour of the considered model. Indeed, as discussed in Section~\ref{sec:model}, the attractor is found to have that dimensionality when the typical length-scales of the dynamics are considered. Results shown in the previous sections, on the contrary, seem to depict a more complex scenario.

 When we apply the stochastic ML analysis described in Section~\ref{sec:machine} using low embedding vector dimension ($n=2$, $n=3$) the results are not so different from those obtained by a more straightforward, physics-inspired analysis as that introduced in Section~\ref{sec:markov}.
  This is quite reasonable, if one considers that the p-vMP approach relies on the determination of the conditional pdf $q(v_{t+1}|x_t,v_t)$, which embodies the same amount of information as the $q(v_{t+1}|x_t,x_{t-1})$ investigated by the ML method. Based on the above conjecture that the attractor of the effective macroscopic dynamics has dimension $\simeq 1.3$, we might expect that such second-order dynamics is enough to catch its main features. Figure~\ref{fig:spettri}(b), which reports the Fourier spectrum of the original and of the reconstructed dynamics for $N=10^5$, clearly shows that this is not the case: indeed the p-vMP procedure and the ML approach with $n=2$ even fail to catch the oscillating behaviour of the dynamics, and they miss the peak at $f\simeq 10^{-2}$.
 
 The reason of this failure can be understood by looking again at Fig.~\ref{fig3}(b). In that case, the histogram of $v_{t+1}$ based on the knowledge of $x_{t}$ and $v_{t}$ is given by the superposition of two peaks with opposite signs, a strong hint that, in that case, such knowledge does not completely identify the actual state of the system. This mismatch is due to the presence of ``deterministic noise'', whose amplitude is larger when $N$ is smaller (see Section~\ref{sec:model}), which brings to an overlap, in the phase space, between two branches of the attractor, characterized by opposite velocities (see Fig.~\ref{fig3}, main plot). Consistently, for values of $N$ large enough ($N\gtrsim10^6$), when the occurrence of cases as the one described in Fig.~\ref{fig3}(b) becomes negligible, the two-variables methods give a fair approximation of the dynamics, and they catch the oscillatory behaviour.
 
 The above scenario seems to reinforce the initial conjecture that the macroscopic dynamics can be actually described by two variables only, and that the failure of our methods in the $N=10^4$ and $N=10^5$ cases is only due to a wrong identification of the state, i.e. to a wrong choice of the two variables to use for the description of the system. However, even when the number of maps is very large (e.g. $N=10^7$), such methods fail to reproduce the low-frequency part of the Fourier spectrum, and we need to switch to a stochastic ML approach with larger $n$ in order to recover sensible results at all scales. Very good results are observed for $n\ge 6$ (see Fig.~\ref{fig:spettri}), so that one might be tempted to conclude that the actual dimension of the attractor is around that value. Bearing in mind the discussion in Section~\ref{sec:tre_n}, it is conversely clear that 3 dimensions should be enough to get a satisfying low-frequency description, once the variables are properly chosen.
 
 An indirect evidence of the validity of this conclusion can be achieved by mean of the \textit{deterministic} ML approach discussed at the end of Section~\ref{sec:machine}. We consider the case $N=10^6$ and $n=10$: as shown in Fig.~\ref{fig:spettri}, this choice provides a very good reconstruction of the dynamics, even if the protocol employs a deterministic approach, because in the large-$N$ limit the stochastic noise is negligible. We can thus expect that this neural network finds a proper way to approximate the true dynamics (whose real microscopic attractor has dimension $10^7$) with a deterministic process involving 10 variables. The GP analysis can then be used to get a hint on the actual dimension of this deterministic dynamics (i.e. the minimum number of variables that would be needed to reproduce it). The results are shown in Fig.~\ref{fig:GPdet}: as expected, at the typical lengthscales of the macroscopic dynamics, we find again that the effective dimension is $\simeq 1.3$; at smaller scales, however, the attractor is found to have a larger correlation dimension, lower than $3$, confirming our previous conjecture built on empirical bases.

By combining all previous observations, we can gain an insightful picture of the multi-scale structure, the memory and dimensionality of the system, at different level of coarse graining, constructed as Markovian processes for the embedding vectors. 
 \begin{itemize}
 \item Delay $n=2$ for $N\geq 10^6$. Approximately one dimensional $(\approx 1.3)$ nearly periodic oscillations, involving a single frequency peak at $f\approx 10^{-2}$
 
 \item Delay $n=6$ for $N\geq 10^5$. This level of coarse graining captures the ``climate'' of the system: low frequencies and main spectrum peaks are fully reconstructed. Minor discrepancies are visible in the high frequency spectrum reconstruction and can be fully appreciated with the cross entropy (cost function). The dimension of the system is approximately 3 and has a main deterministic contribution for $N\geq 10^6$.
 
 \item Delay $\tilde n\approx 10$. Both ``climate'' and ``weather'' are reconstructed. To our available computational power, this level of coarse graining allows the best reconstruction of both low and high frequencies (as measured with cross entropy cost function). This works also for the ``problematic'' case $N=10^4$, which lacks a clear deterministic backbone in the attractor.
 
 \item Delay $n\geq \tilde{n}\approx 10$. Residual effect of the high dimensional deterministic dynamics. Information cannot be effectively extracted from it and should be modelled as noise in a coarse grained description. It becomes less relevant for $N\geq 10^6$.  From machine learning we can estimate the Shannon entropy rate of this unresolved fast dynamics.
 
 \end{itemize}
 This non-trivial structure of the macroscopic dynamics was not evident from either equation or data.



\section{\label{sec:conclusions} Conclusions}

The study of macroscopic dynamics from data in systems with many degrees of freedom is a challenging problem in modern physics, and it has a wide range of applications, from turbulence to climate physics.  In this paper we have employed a ML approach to investigate the non-trivial macroscopic dynamics observed in a system of coupled chaotic maps; the outcomes have been compared to those of a more straightforward, physics-inspired analysis. Our aim was twofold: on the one hand, we wanted to test the ability of a stochastic ML approach in reproducing such non-trivial dynamics; on the other hand, we aimed at understanding whether the ML analysis could shed some light on the physics of the original model, for instance by determining the number of dimensions of its macroscopic dynamics.

Let us notice that the two purposes of understanding and forecasting are strongly interwoven.
Predictions either fail or succeed: in the former case, it is hard to reach confident conclusions, since the failure may be either due to poor variable choice or to an inefficient machine learning approach; if predictions are successful, though, we can state that certain variables or information are \textit{sufficient} (not necessary though) for building a model with a given degree of coarse graining. We can even control the level coarse-graining by employing different metrics.
While indeed many powerful black-box tools exist to tackle forecasting problems, our choice of a direct technique allows for great control and easy manipulation of what variables and information (in our case, the delay vector) the network is using in building a certain model.

Our analysis shows that the stochastic ML approach introduced in Section~\ref{sec:machine} can reproduce the dynamics with remarkable accuracy: when the embedding vector passed to the neural network has dimension $n=2$, the results are comparable with those obtained by a direct reconstruction of the stochastic position-velocity dynamics, as outlined in Section~\ref{sec:markov}; by increasing the length of the delay vector, the quality of the reconstruction gets even better, until the Fourier spectrum of the original and of the reproduced dynamics can be barely distinguished. This is indeed remarkable, since in the training process no direct information about the low-frequency behaviour of the dynamics is provided to the neural network.

Our systematic analysis also gives some hint about the dimensionality of the macroscopic dynamics. The value of $n$ at which the reconstruction starts to be reliable at every scale can be seen as an upper bound to the dimension of the attractor of the macroscopic dynamics. A better job can be done by investigating whether \textit{all} elements of the $n$-dimensional embedding vector are actually relevant to the reconstruction. This task may be accomplished by a careful analysis of the cross entropy as a function of $n$: larger decreases are expected to be related to ``more relevant'' elements of the embedding vector. In this sense, the proposed procedure is not only able to determine an upper bound to the quantity of variables that are actually needed to describe the original trajectory. The idea of ``compressing'' the signal and only keep relevant information is not new in machine learning (and specific machine learning approach to attractor size estimation can be found in literature~\cite{linot2020deep}) but, unlike some automatic dimensional-reduction approaches (e.g. autoencoder neural networks, see Appendix~\ref{app:ML}), the technique employed in this paper also provides some direct hint on \textit{which} variables may be expected to be relevant in modelling the macroscopic dynamics.

We have thus shown that our approach can extract information ranging from multi-scale structure, memory effects, entropy rates, effective dimension and even variable selection. All these results could have probably be obtained with carefully designed standard numerical observables; the advantage of using this machine learning-based, modelling-oriented, observables - even for preliminary investigations  - is that, by construction, they are relevant in a forecasting framework, a feature which is not obvious in general.
As a side note, let us also stress that once we are confident enough about the physical reconstruction, the network-based model can even be used as a data augmentation. In our case, we could produce a very long deterministic trajectory ($T=10^7$) to run Grassberger-Procaccia algorithm, something that would have been computationally critical, e.g., for $N=10^6$.
 In this perspective, an exploration-oriented machine-learning approach may probably be optimal when used alongside with standard physical techniques and powerful black-boxes tools, for maximizing both insight and performance.

\begin{acknowledgments}
We wish to acknowledge useful discussions with M.Cencini and A.Vulpiani.\\
M.B. acknowledges the financial support of MIUR-PRIN2017 \textit{Coarse-grained description for non-equilibrium systems and transport phenomena (CO-NEST)}.
\end{acknowledgments}

\appendix

\section{Two-band structure in model~\eqref{eq:dyn} }
\label{sec:appendixA}
In this Appendix we discuss some properties of model~\eqref{eq:dyn}, providing some hint about
the qualitative behaviour discussed in Section~\ref{sec:model}: even if an exhaustive analysis of the problem
is out of the scope of this paper, some indications may be useful to understand the origin of the main features shown by the considered dynamics.
\begin{figure}
\includegraphics[width=0.99\linewidth]{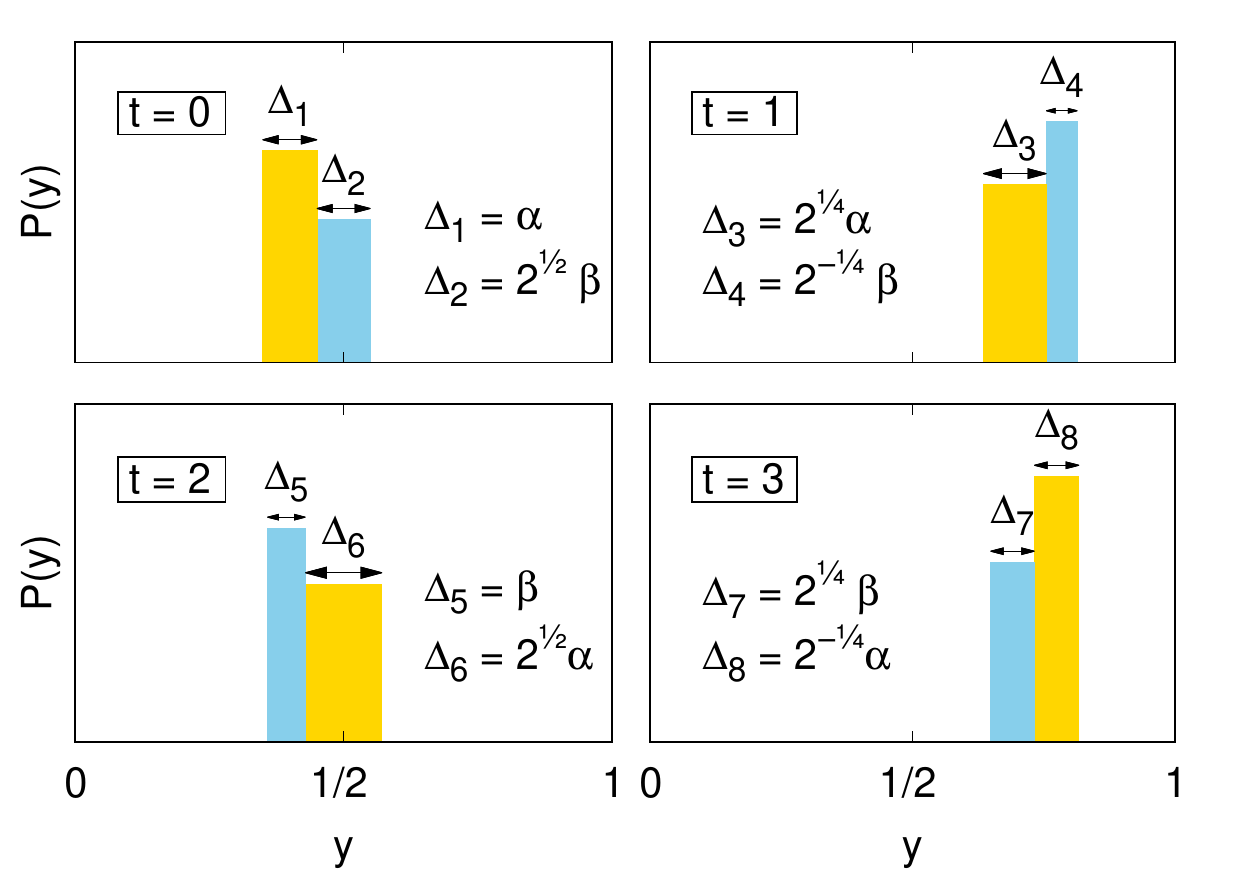}
 \caption{\label{figA1} Scheme of the 4-steps dynamics which can be realized by model~\eqref{eq:dyn} if condition~\eqref{eq:parmagici} is satisfied. Each panel represents the distribution of the maps at a given time.}
\end{figure}

First, one has to recognize that the quantity $a(1-\varepsilon)$ is crucial for the dynamics, as it determines the ``expansion rate'' of the distance between two maps $y^{(i)}_t$ and $y^{(j)}_t$, if at time $t$ they assume values that are both larger or both smaller than 0.5. As a consequence, 
$a(1-\varepsilon)$ has a major role in the evolution of the distribution $p(y_t)$ of the maps.
Interesting situations are found when $a(1-\varepsilon)=2^{1/n}$, with $n$ integer, since in those cases it is usually possible to design piece-wise constant distributions which are mapped into themselves after $n$ time steps. A trivial example is given by the distribution
\begin{equation}
 p(y)=
 \begin{cases}
 1/\alpha \quad \quad \text{if} \quad \frac{1-\alpha}{2} < y < \frac{1+\alpha}{2}\\
 0\quad \quad\text{otherwise}\,,
 \end{cases}
\end{equation} 
which is mapped into itself after 2 steps if $a(1-\varepsilon)=\sqrt{2}$ and $\alpha=2-4a^{-1}+2a^{-2}$.

A less trivial case is represented by the case 
\begin{equation}
\label{eq:parmagici}
a(1-\varepsilon)=2^{1/4}\,.
\end{equation}
Let us consider the distribution
\begin{equation}
 p(y)=
 \begin{cases}
 \gamma/\alpha \quad \quad \text{if} \quad \frac{1}{2}-\alpha-\frac{\beta}{\sqrt{2}} < y < \frac{1}{2}-\frac{\beta}{\sqrt{2}}\\
 (1-\gamma)/\alpha \quad \quad \text{if} \quad  \frac{1}{2}-\frac{\beta}{\sqrt{2}} < y < \frac{1}{2}+\frac{\beta}{\sqrt{2}}\\
 0\quad \quad\text{otherwise}\,,
 \end{cases}
\end{equation} 
which is sketched in the first panel of Fig.~\ref{figA1}. Here $\gamma$ is a 
parameter between 0 and 1 which sets the relative areas of the two rectangular 
``bands'' of $p(y)$, whose widths are determined by the constant parameters 
$\alpha$ and $\beta$. For every couple of parameters $(a,\varepsilon)$ 
satisfying the relation~\eqref{eq:parmagici}, it is possible to find an infinite 
number of triplets $(\alpha, \beta, \gamma)$ such that, if the system is 
distributed according to $p(y)$ at time $t=0$, it will assume the same 
distribution at $t=4$, following the steps sketched in Fig.~\ref{figA1}. 
Basically, one has to impose the condition that at time $t=2$ and $t=4$ the 
center of the right band is placed at $y=1/2$, so that all rectangular bands are 
mapped into new rectangles by the dynamics. Since this condition results in two 
constraints, and the distribution is characterized by 3 parameters, we can find 
an arbitrary number of different $p(y)$ showing the 4-steps recurrency shown in 
Fig.~\ref{figA1}.

The above discussion should clarify some aspects of the dynamics of 
model~\eqref{eq:dyn} when the parameters are chosen according to 
Eq.~\eqref{eq:parameters}. Indeed, $a(1-\varepsilon)=1.19$ is very 
close to condition~\eqref{eq:parmagici}, since $2^{1/4}\simeq1.1892$. Although 
the highly-nonlinear character of the dynamics hinders the possibility of a 
simple perturbative approach to the problem of modelling $z_t$, at least the 
basic 4-steps temporal structure observed in numerical simulations seems to be 
closely connected to the peculiar case~\eqref{eq:parmagici}. The oscillatory 
behaviour on slower scales, shown by the system for $N$ sufficiently large (see 
Fig.~\ref{fig1}), might be interpreted as a ``wandering'' of the 
system among the eigenstates of the 4-steps ``unperturbed'' dynamics satisfying 
condition~\eqref{eq:parmagici}.

\section{Minimal introduction to feed-forward neural networks}\label{app:ML}

Here, for sake of self-consistency, we provide a minimal description of a simple fully connected feedforward neural network (FFNN) adapted to the procedure we have followed. We refer to literature for details and for
any further technical or theoretical explaination, see for instance~\cite{scarselli1998universal,hertz2018introduction,mehlig2019artificial,jentzen2020overall,nishimori2001statistical,du2013neural}. A FFNN is a map
\begin{equation}
F:\mathbb{R}^{n_{0}}\to \mathbb{R}^{n_{N}}
\end{equation}
where $n_0$ is the dimension of the input and $n_N$ is the dimension of the output. The network can be described as the composition of layers:
\begin{equation}
F=F^{(N)}\circ F^{(N-1)}\circ ... \circ F^{(1)}
\end{equation}
with
\begin{align}
F^{(i)}:\mathbb{R}^{n_{i-1}}\to \mathbb{R}^{n_{i}},
\end{align}
and
\begin{equation}
F^{(i)}_a(\boldsymbol{x})=\sigma_i\left(\sum_{b=1}^{n_{i-1}} w^{(i)}_{ab}x_b+\theta_a^{(i)}\right).
\end{equation}
The matrix-array tuple $\Theta^{(i)}=(w^{(i)},\theta^{(i)})$ describes the $i$ layer along with the activation function $\sigma_i$, which is, in our case, the ReLU (Rectified Linear Unit) function
\begin{equation}
\sigma (x)=\begin{dcases}
x\mbox{ if } x>0\\
0\mbox{ if } x\leq 0,
\end{dcases}
\end{equation}
for all but the last layer. In this work, the last $\sigma_i$ is either the identity (Linear Unit), for the deterministic network, or the soft-max, for the stochastic network. Therefore, $F=S$ in equation~\eqref{Sn_} with
\begin{equation}\label{eqSn_app}
F_a=\sum_{b=1}^{n_{N-1}} w^{(N)}_{ab} F^{(N-1)}_b+\theta^{(N)}_a
\end{equation}
and $F=Q$ in equation~\eqref{Qn_} 
\begin{equation}
F_a=\frac{\exp(\sum_{b=1}^{n_{N-1}} w^{(N)}_{ab} F^{(N-1)}_b+\theta^{(N)}_a)}{\sum_{c=1}^{n_N} \exp(\sum_{b=1}^{n_{N-1}} w^{(N)}_{cb} F^{(N-1)}_b+\theta^{(N)}_c)}
\end{equation}
respectively.

A fundamental result of machine learning is that any smooth function on a compact can be approximated arbitrarily well within the class of functions given above~\cite{scarselli1998universal}. Therefore, for any fitting problem, some $N$, a set $\{n_i\}_{i=1}^N$ and a set of parameters $\Theta=\{\Theta^{(i)}\}_{i=1}^N$ exist such that the problem can be solved with any fixed precision. 

A cost function $C(\Omega)$ measures how well a network can perform a certain task on the set $\Omega$: lower costs are associated to better performances. For instance, in a classification problem, the cost function can be the fraction of misclassified items or the mean square error when fitting a function. For any given cost function $C$, training set $\Omega_{train}$, $N$, $\{n_i\}_{i=1}^N$ and $\{\sigma_i\}_{i=1}^N$, one can find the optimal parameters
\begin{equation}
\Theta^*\approx\arg\min_{\Theta} C(\Omega_{train})
\end{equation}
by using one of the many robust and efficient variants of the standard gradient descent algorithm (GD). We have employed the ADAM~\cite{kingma2014adam} variant. 

The performance of machine learning is not fully explained by its remarkable fitting capability: a number of mathematical results from statistical theory of learning show that, when properly trained, a suitable network has a good performance in processing data it has never seen before; when this happens, the network is capable of \textit{generalizing}~\cite{du2013neural}. Consistently, in order to check if the network has been suitably trained, we need a set of independent (w.r.t. the training set) data - the validation/test set $\Omega_{test}$ - to validate the generalization performance, measured as $C(\Omega_{test})$ . Indeed, while minimising $C(\Omega_{train})$ is the mean, minimising $C(\Omega_{test})$ is the goal of learning.

\bibliography{biblio}

\end{document}